\documentclass{aa}
\usepackage[table]{xcolor}

\usepackage{soul}%

\usepackage{natbib}
\usepackage{graphicx}
\usepackage{color}
\usepackage{txfonts}
\usepackage{bm} 

\graphicspath{{./}{figs-ps/}{figs/}}

\begin{document}

   \title{Emission of solar chromospheric and transition region features related to the underlying magnetic field}

\titlerunning{Solar chromospheric and TR features related to underlying magnetic field}

   \author{K.~Barczynski\inst{1}
          \and  H.~Peter\inst{1}
          \and  L.~P.~Chitta\inst{1}
          \and  S.~K.~Solanki\inst{1,2}
          }

   \institute{Max Planck Institute for Solar System Research, 
             37077 G\"ottingen, 
             Germany, 
             email: barczynski@mps.mpg.de
           \and
             School of Space Research, 
             Kyung Hee University, 
             Yongin, Gyeonggi, 446-701, 
             Republic of Korea}

   \date{Received ...; accepted...}

 
  \abstract
   {The emission of the upper atmosphere of the Sun is closely related to magnetic field concentrations at the solar surface.}
   {It is well established that this relation between chromospheric emission and magnetic field is nonlinear. Here we investigate systematically how this relation, characterised by the exponent of a power-law fit, changes through the atmosphere, from the upper photosphere through the temperature minimum region and chromosphere to the transition region.
}
   {We used spectral maps from the Interface Region Imaging Spectrograph (IRIS)  \ion{covering Mg}{ii} and its wings, \ion{C}{ii,} and \ion{Si}{iv} together with magnetograms and UV continuum images from the Solar Dynamics Observatory. After a careful alignment of the data we performed a power-law fit for the relation between each pair of observables and determine the power-law index (or exponent) for these. This was done for different spatial resolutions and different features on the Sun.}
   {While the correlation between emission and magnetic field drops monotonically with temperature, the power-law index shows a hockey-stick-type variation: from the upper photosphere to the temperature-minimum  it drops sharply and then increases through the chromosphere into the transition region. This is even seen through the features of the \ion{Mg}{ii} line, this is, from k1 to k2 and k3.
It is irrespective of spatial resolution or whether we investigate active regions, plage areas, quiet Sun, or
coronal holes.}
   {In accordance with the general picture of flux-flux relations from the chromosphere to the corona, above the temperature minimum the sensitivity of the emission to the plasma heating increases with temperature. Below the temperature minimum a different mechanism has to govern the opposite trend of the power-law index with temperature. We suggest four possibilities, in other words, a geometric effect of expanding flux tubes filling the available chromospheric volume, the height of formation of the emitted radiation, the dependence on wavelength of the intensity-temperature relationship, and the dependence of the heating of flux tubes on the magnetic flux density.
   }
 
 {}

   \keywords{Sun: chromosphere 
         --- Sun: transition region
         --- magnetic fields 
         --- Sun: UV radiation 
         --- Sun: activity
         }

 \maketitle

%

\begin{table*}[ht]
\caption{IRIS data sets used in this study. 
\label{table:t1}}
\centering                          
\begin{tabular}{l@{~}l c c l c c c r}        
\hline\hline                 
   & &  date & start (UT) & end (UT)  & exposure\,time & raster type\tablefootmark{(a)} & X\tablefootmark{(b)} & Y\tablefootmark{(b)} \\    
\hline                        
active region & (AR 11850) &  24 Sep. 2013 & 11:44:43 & 12:04:10 & ~\,2\,s  &  & $-$265\arcsec & $+$88\arcsec \\      
quiet Sun & (QS) &  13 Oct. 2013 & 23:27:28 & 02:59:15\,(+1d) & 30\,s  & 400x0.35\arcsec & $-$120\arcsec & $-$41\arcsec \\
coronal hole & (CH) &  12 Oct. 2013 & 12:20:28 & 15:52:15 & 30\,s  &  & $+$314\arcsec & $-$140\arcsec \\

\hline                                   
\end{tabular}
\tablefoot{%
\tablefoottext{a}{In all cases large dense rasters were performed with a field-of-view of about 140\arcsec$\times$170\arcsec. These required 400 raster steps of 0.35\arcsec\ each in the east-west (solar-X) direction.}
\tablefoottext{b}{The values for X and Y give the distance of the centre of the field-of-view from disc centre in the solar-X and Y directions.}
}

\end{table*}

\section{Introduction}\label{S:introduction}


One of the major proxies to characterise the
magnetic activity of a (cool) star is its chromospheric emission \cite[e.g.][]{2008LRSP....5....2H,2012LRSP....9....1R}.  In particular, the \ion{Ca}{ii} H and K lines in the ultraviolet (UV) at 3968\,\AA\ and 3934\,\AA\ and the \ion{Mg}{ii} h and k lines further in the UV near 2804\,\AA\ and 2796\,\AA\ are used. Using an index based on the \ion{Ca}{ii} lines a longterm monitoring programme was set up at the Mount Wilson Observatory to investigate stellar activity cycles \cite[][]{1978ApJ...226..379W,1995ApJ...438..269B}. To interpret the chromospheric indices in terms of magnetic activity and cyclic variation solar observations are a key. This is because only on the Sun can we directly study the spatio-temporal relation between chromospheric emission and the underlying magnetic field. The Interface Region Imaging Spectrograph \cite[IRIS;][]{2014SoPh..289.2733D} provides new solar data to this study of the spatial relationship of the chromospheric emission and the magnetic field through spectral maps in the \ion{Mg}{ii} lines. And IRIS adds information from the photosphere (through the wings of \ion{Mg}{ii}) and in particular from the transition region (from the doublets of \ion{C}{ii} and \ion{Si}{iv} around 1335\,\AA\ and 1400\,\AA).

Investigations of the relation of chromospheric emission and the (unsigned) photospheric magnetic flux density $|B|$ (interchangeable used as magnetic field hereafter) on the Sun date back more than half a century and are mostly based on studies of the H and K lines of \ion{Ca}{ii}. Comparing spectroheliograms and magnetograms in plage areas \citet{1959ApJ...130..366L} found a spatial correspondence between |$B$| and \ion{Ca}{ii}. The first quantitative studies for the quiet Sun revealed a linear dependance \citep{1975ApJ...200..747S,
1998SoPh..179..253N}. However, \citet{1989ApJ...337..964S} suggested a power-law relation for the plage area, with an exponent of about 0.6, that is, with a much weaker dependence of the chromospheric emission on the magnetic field.  Likewise, later studies argued that a power-law (with an exponent less than one)   is found in plage and quiet Sun network regions \cite[e.g.][]{1999ApJ...515..812H,2005MmSAI..76.1018O,2007A&A...466.1131R}. A more detailed analysis of the quiet Sun was done by \citet{2009A&A...497..273L}. They 
compared magnetograms and \ion{Ca}{ii} K filtergrams  separately for the network and internetwork regions of the quiet Sun. While for the network they found the well established power-law relation (exponent of about 0.5), in the internetwork \ion{the Ca}{ii} K-line emission is independent of the underlying magnetic field. This is consistent with the general idea that the internetwork is not dominated by magnetic fields.

 
While there is a clear relation between the magnetic field and the chromospheric emission, even for the lowest magnetic field values the emission is not zero. The minimum energy flux through radiation from the chromosphere (and higher layers in the atmosphere) in the absence of magnetic field is termed basal flux \citep{1989ApJ...337..964S, 1992A&A...258..507S}. This basal flux found in the nonmagnetic regions is generally thought to be due to acoustic heating  \cite[e.g.][]{1991SoPh..134...15R}. Consequently, when searching for a physical connection between the surface magnetic field and the chromospheric emission one has to subtract this basal flux from the radiation from the chromosphere (or from  higher and hotter regions).

A direct quantitative comparison of results from previous observational studies on the relation of the magnetic field to the chromospheric emission is difficult. Naturally the authors used different instrumentation, for example, with different sensitivity with respect to the magnetic field or various widths of the spectral bandpass (or resolution) for the chromospheric emission. However, the common property of all these observations is a power-law relation between magnetic field and chromospheric emission with a power law index below unity.

This relation has been interpreted already by \cite{1989ApJ...337..964S} as being due to the geometry of the magnetic field expanding from the concentrations in the photosphere into the chromosphere. A higher (average) magnetic field strength corresponds to a denser packing of the (wine-glass shaped) magnetic flux tubes. Once these expanding tubes meet in the chromosphere (where \ion{Ca}{ii} forms) even a denser packing of the flux tubes cannot increase the chromospheric emission. Basically with increasing magnetic field the chromospheric emission saturates. Instead of a linear relation between field and emission the relation gets flatter for higher field strength, corresponding to a relation with a power-law exponent smaller than one. This idea was later confirmed by a proper two-dimensional model \cite[][]{1991A&A...250..220S}.

When comparing the magnetic field and the emission from hotter regions of the upper atmosphere, the picture changes. For example \citep{2003ApJ...598.1387P} found the power-law exponent to be (slightly) larger than one, and when considering the whole solar disc averages they found values being as high as almost two. We do not further discuss the implications of these results on coronal heating here, but highlight the question on how the relation of the magnetic field to the emission changes through the different temperature regimes of the atmosphere. In this study we will concentrate on the change from the upper photosphere through the temperature minimum region and the chromosphere to the transition region. This will then provide a test to the geometric picture explaining the relation of magnetic field to emission by investigating how this relation might change with temperature based on the model idea. This study also provides a first quantitative analysis of the intensity from the transition region vs. |$B$| and fills the gap between previous studies that focused mostly on the chromosphere and the corona.

In this study we will use the term mag-flux relations for the relation of the magnetic field in the photosphere to the radiative flux from the different regions of the atmosphere. In addition to this we will also investigate the mutual relations of the radiative fluxes and term these flux-flux relations (in accordance with the existing literature). The flux-flux relations have been investigated extensively in the framework of stellar studies, simply because of the frequent lack of knowledge about the surface magnetic field of the stars under consideration \cite[e.g.][]{1981ApJ...247..545A,1986A&A...154..185O}. Here we investigate the flux-flux relations mainly because we can analyse emission features from different  temperature regimes observed with the same spectrograph simultaneously through the same slit. Thus any problems of spatial misalignment can be ruled out.

To get a good continuous temperature coverage we use data from IRIS (cf.\ Sect.\ref{S:observation-observation}).
In particular the \ion{Mg}{II} lines provide good diagnostics throughout the chromosphere \cite[][]{2013ApJ...772...89L,2013ApJ...772...90L,2013ApJ...778..143P}. Using the wings of \ion{Mg}{II} we add information from the upper photosphere, and the \ion{C}{ii} and \ion{Si}{iv} provide details on the transition region.
We complement these data with magnetograms from the Helioseismic and Magnetic Imager~\citep[HMI;][]{2012SoPh..275..207S} to study the mag-flux relations.

After a discussion on the  preparation of the data and the analysis method (Sects.\,\ref{S:observation} and \ref{S:method})
we show in Sect.\,\ref{S:results} that there is a continuous variation of the power-law exponent of the mag-flux relation, all the way from the photosphere into the transition region. 
Based on these results we discuss in Sect.\,\ref{S:discussion}
the drop of the power-law index from the photosphere to the chromosphere in terms of four possible scenarios.


\section{Observations and preparation of data}\label{S:observation}
\subsection{Observations}\label{S:observation-observation}

In this work we concentrate on observations acquired simultaneously by the Interface Region Imaging Spectrograph \citep[IRIS,][]{2014SoPh..289.2733D} and the Solar Dynamics Observatory \citep[SDO,][]{2012SoPh..275....3P}.
We focus on the active region AR11850 with an extended plage area and compare the derived properties with data sets covering a quiet Sun region and an (on-disc) coronal hole (cf.\ Table\,\ref{table:t1}).

In our study we use IRIS data covering the chromosphere and transition region using simultaneously recorded spectra and slit-jaw-images.
The large dense rasters from IRIS cover 400 steps with a step size of 0.35\arcsec\ providing maps of the spectral line properties with a full field-of-view of about 140\arcsec$\times$170\arcsec. The spatial scale along the slit is about 0.17\arcsec/pixel.
Here we concentrate on \ion{Mg}{ii}\,k, \ion{C}{ii} and \ion{Si}{iv}.
While the different features of the \ion{Mg}{ii} profile, such as, the k3 self-reversal, the k2r peak and the k1r minimum, originate from the upper to the lower chromosphere, \ion{C}{ii} and \ion{Si}{iv} stem from the transition region.
Details of the emission line properties are summarised in Table\,\ref{table:t2}.
The temperature of lines' formation should be taken with a grain of salt because these would apply only under equilibrium conditions.
On the real Sun the atmosphere is quite dynamic and the traditional view of a stable stratified atmosphere certainly does not apply.
Still the formation temperatures listed in Table\,\ref{table:t2} provide some ordering of the lines (and their spectral features) with temperature in the upper solar atmosphere, in an average sense.

\begin{table*}[ht]
\caption{Overview of the observations.} \label{table:t2}
\centering
\begin{tabular}{l@{~~}l l c c  c}
\hline\hline
   \multicolumn{2}{c}{instrument} 
&  $\lambda$\,[\AA]\tablefootmark{(a)} 
&  line / feature 
&  ${\log}\,T\,[{\rm{K}}]$\tablefootmark{(d)} 
&  atmospheric regime\tablefootmark{(e)}\\ 
\hline
  & HMI & 6173 &  magnetogram\tablefootmark{(b)} &  -- &  photosphere \\
\raisebox{1ex}[-1ex]{SDO}     
  & AIA & 1600 &  continuum\tablefootmark{(c)} & <3.6~~ & upper photosphere / $T\!$-min\tablefootmark{(f)}\\
\hline 
 & & 2796.9 &  \ion{Mg}{ii} k1r &  3.6 & $T\!$-min / lower chromosphere\\
 & & 2796.6 &  \ion{Mg}{ii} k2r &  3.8 & middle chromosphere \\
\multicolumn{2}{c}{IRIS} & 2796.4 &  \ion{Mg}{ii} k3 & 3.9 & upper chromosphere \\
 & & 1335.7 &  \ion{C}{ii}  & 4.6 & low transition region \\
 & & 1393.8 &  \ion{Si}{iv} & 4.9 & transition region \\
\hline                      
\end{tabular}
\tablefoot{%
\tablefoottext{a}{The AIA 1600\,\AA\ images are acquired in a roughly 50\,\AA\ wide wavelength band. For lines observed with IRIS the wavelength of the respective features of the \ion{Mg}{ii}\,k line or the rest wavelength of \ion{C}{ii} and \ion{Si}{iv} are given (see Sect.\,\ref{S:observation-dpi}).}
\tablefoottext{b}{In active regions HMI can also provide the full vector of the magnetic field. We use the line-of-sight component of the magnetic field from HMI.}
\tablefoottext{c}{This 1600\,\AA\ band contains also the \ion{C}{iv} doublet at 1548\,\AA\ and 1550\,\AA. Except for small patches of strongly enhanced emission around 10$^5$\,K this band does not show transition region structures on the disc (e.g. compare panels b and d in Fig.\,\ref{fig_a}).}
\tablefoottext{d}{The formation temperatures for the 1600\,\AA\ continuum and the \ion{Mg}{ii}\,k line features are taken from \citet{1981ApJS...45..635V}, the values for \ion{C}{ii} and \ion{Si}{iv} are taken from \citet{2006ApJ...638.1086P}. } 
\tablefoottext{e}{This ordering should be a guideline only, in particular when considering the dynamic nature of the solar upper atmosphere.} 
\tablefoottext{f}{$T\!$-min denotes the temperature minimum.} 
}
\end{table*}

To investigate the response of the upper atmosphere to the photospheric magnetic field we use line-of-sight magnetograms from the Helioseismic and Magnetic Imager~\citep[SDO/HMI;][]{2012SoPh..275..207S}.
These provide information on the full solar disc with a plate scale of 0.5\arcsec/pixel at a cadence of 45\,s.
For a reliable alignment between the IRIS raster maps and the HMI magnetograms (Sect.\,\ref{S:SDO.prep}) and to investigate the temperature minimum region we employ the 1600\,\AA\ channel of the Atmospheric Imaging Assembly~\citep[SDO/AIA;][]{2012SoPh..275...17L}.
The 1600\AA\ images of AIA provide data of the full solar disc with a 0.6\arcsec/pixel plate scale at a temporal cadence of 24\,s.

 \begin{figure}
   \centering
   \includegraphics[width=8.8 cm]{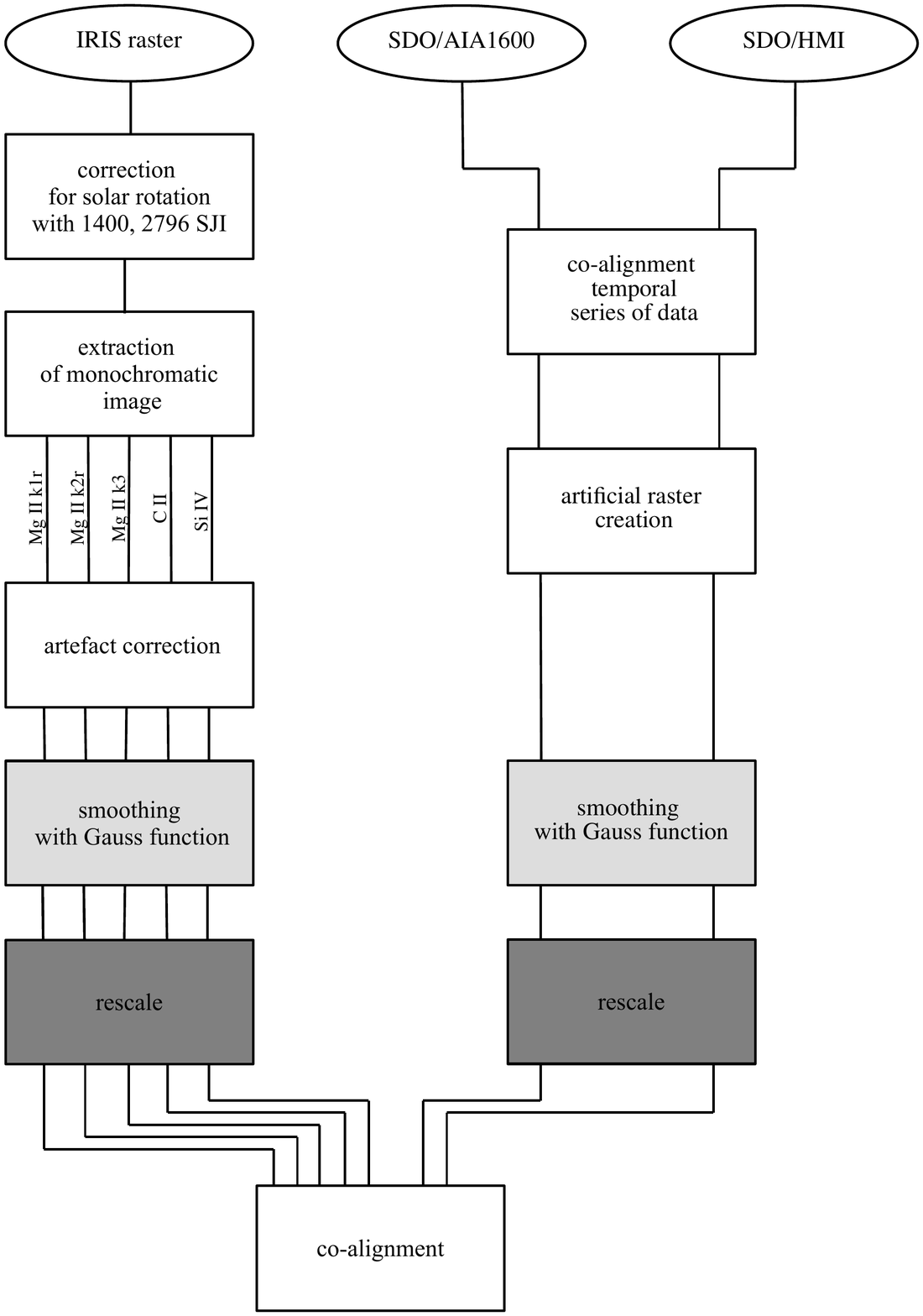}
      \caption{Flowchart of data preparation process. See Sect.\,\ref{S:observation-dp}.}
              \label{fig_b}
    \end{figure}

\subsection{Data reduction}\label{S:observation-dp}

\subsubsection{IRIS spectral maps}\label{S:observation-dpi}

After standard procedures to correct the IRIS level-2 data for dark current, flat field and geometric distortions, we apply a five-step preparation procedure. This is shown schematically in Fig.\,\ref{fig_b}.

First we correct the IRIS raster data for solar rotation.
The next and main step is the production of monochromatic maps to obtain spectrally pure images of the line profile features.
In the case of \ion{Mg}{ii} we locate the position of the self-reversal, k3, and the maximum of the red wing, k2r, and the emission at the k3 minimum and the k2r peak.
For this we employ the IRIS reduction software available in SolarSoft\footnote{iris\_get\_mg\_features\_lev2.pro available at SolarSoft, \url{http://www.lmsal.com/solarsoft/}.}.  
For the k1r feature, the minimum in the red wing of the \ion{Mg}{ii}, we use a 0.7\,\AA\ wide window around the expected wavelength of k1r (cf.\,Table\,\ref{table:t2}), apply a spline interpolation, and calculate the minimum intensity.
The optically thin line of \ion{Si}{iv} shows a single-peaked spectrum almost everywhere \citep[for exceptions see e.g.][]{2014Sci...346C.315P}, and we apply a spline interpolation to calculate the peak intensity.
The line of \ion{C}{ii} is not optically thin and shows signatures of a self-reversal, in particular in plage-like regions.
Therefore we do not use the peak intensity but calculate the total line intensity (integrated over the line after subtraction of the continuum).
Still, if one would use the peak intensity for \ion{C}{ii} the results remain basically unchanged (because even at the high spectral resolution of IRIS the line appears single peaked in a large part of the field-of-view, in particular in more quiet regions at low signal-to-noise).

After the extraction of the emission of the line-profile features we corrected for artefacts.
Mainly these are rows or columns of bad data (missing data or obvious problematic count rates).
These data points were replaced with the interpolated data from adjacent pixels. The IRIS spectral maps are easily aligned spatially among themselves through the fiducial mark on the slit.

We want to investigate the data at different spatial resolutions to study the effect of the resolution on the relation of the upper atmosphere emission to the magnetic field.
For this we convolve the data with a Gaussian to reduce the spatial resolution and finally bin the smoothed data to the required resolution.
Here we use plate scales of 0.5, 1.5, 3.0 and 6.0\arcsec/pixel.

From this procedure we obtain five rotation-corrected, monochromatic, artefact free and rigidly aligned IRIS images for each of the spatial scales for each of the regions listed in Table\,\ref{table:t1}.

\subsubsection{SDO imaging and surface magnetic field}\label{S:SDO.prep}

For the SDO data we employ a four-step procedure that is also described schematically in Fig.\,\ref{fig_b}.
Below we describe the procedure for the AIA 1600\,\AA\ data, but the HMI data are prepared in the same way.

First we extract the time series of the AIA data for the whole time of the raster scan and align this time series.
The main step is then to extract the AIA data at the position of the IRIS slit in the AIA image closest in time.
Through this we create a raster map of AIA data that is co-temporal to the IRIS raster maps.
This step is crucial to account for the changes in the AIA\,1600\,\AA\ channel (and the HMI data) during the comparably long time the raster maps are acquired (cf.\ Table\,\ref{table:t1}).
This ensures to have SDO data that are co-spatial and co-temporal with the IRIS maps --- prerequisite to get a reliable relation between the data products of the different instruments.
Just like for the IRIS data we create SDO data sets at four spatial scales (0.5, 1.5, 3.0 and 6.0\arcsec/pixel), which means that, first convolving with a Gaussian and then re-binning. 

In the final step combining the IRIS and SDO data we align the IRIS maps with the SDO data.
For this we first align the \ion{Mg}{ii}\,k1r maps with the AIA 1600\AA\ artificial raster maps, and these to the HMI magnetograms (more precisely to the artificial raster map of the HMI magnetograms).

We use HMI line-of-sight magnetograms and not the vertical component derived from the full vector magnetograms, because the goal of this study is to relate the intensity of small-scale features to the underlying magnetic field. Because the small scale features change quickly, on the time scale of a few minutes, we have to ensure that the intensity maps acquired through the spectrograph slit and the magnetograms are co-temporal.
For this we create the raster maps from the HMI magnetogram time series, in the same way as outlined above for the AIA 1600 \,\AA\ channel. 
The line-of-sight magnetograms provide a time cadence of 45\,s which is sufficient for this purpose.
In contrast, the cadence of the full vector magnetic field data is only 720 s or 12 minutes, which is too slow.
To minimise the effect of not using the true vertical component, the regions we chose to investigate are quite close to disc centre. Their $\mu$ angles are 0.96, 0.99, and 0.93, respectively.
Therefore the line-of-sight magnetic field will be close to the vertical component.

\begin{figure*}
   \centering
   \includegraphics[width=17.6 cm]{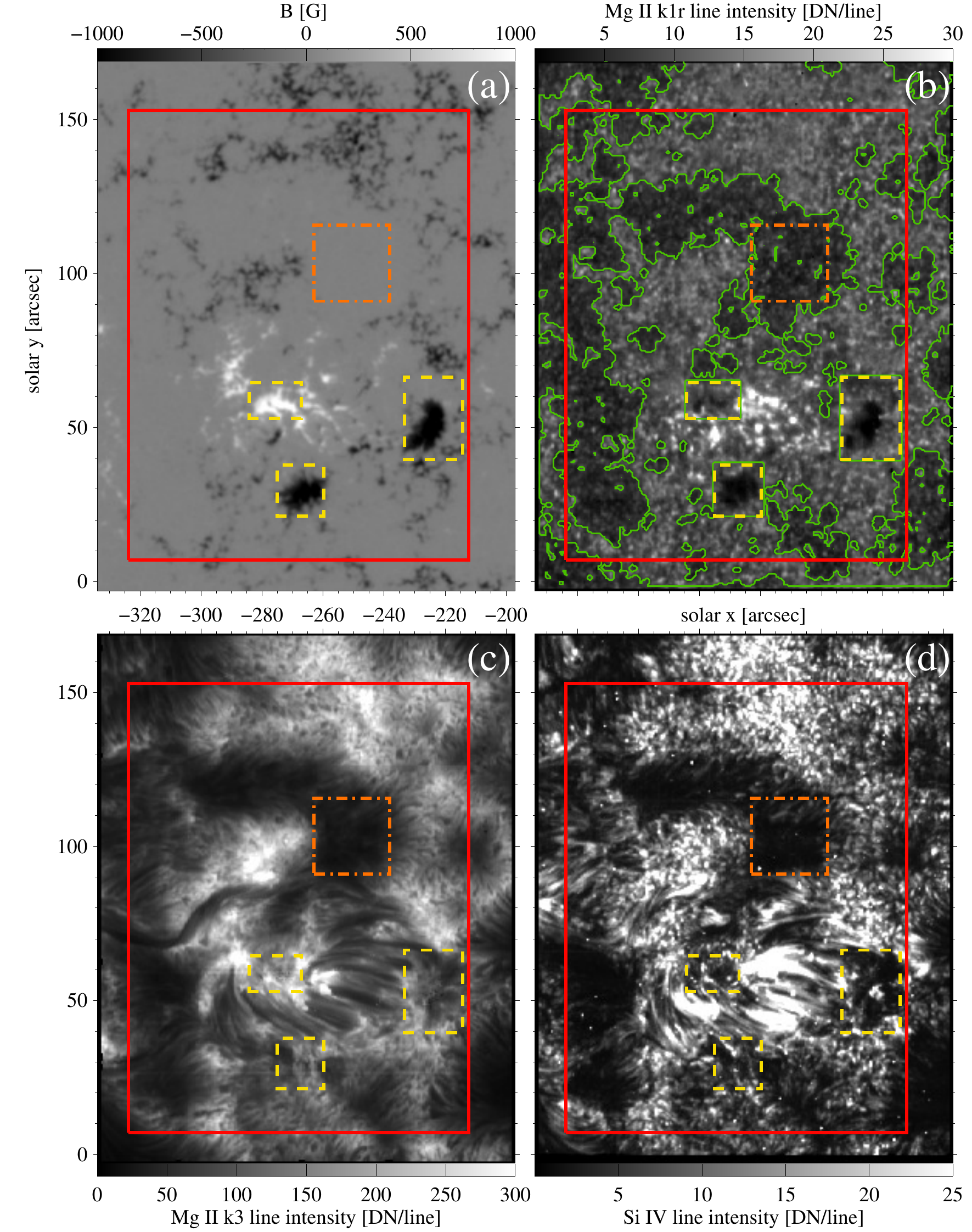}
   \caption{Active region from IRIS raster scan and magnetic field context from HMI.
The images display the active region 11850 (cf.\,Table\,\ref{table:t1}). %
The panels show (a) the SDO/HMI line-of-sight magnetogram and the raster maps by IRIS in (b) \ion{Mg}{ii}\,k1r, (c) \ion{Mg}{ii}\,k3, and (d) \ion{Si}{iv}.
The lines and contours highlight the regions used to define the active region and plage areas (see Sect.\,\ref{S:roi}).
The three yellow rectangles indicate the location of pores and sunspots, the large red rectangle the full extent of the active region.
The green contours (only in panel b) show the location of the plage area. The threshold for the definition of the plage area was calculated in the orange rectangle indicating a quiet region.} 
    \label{fig_a}
    \end{figure*}

In summary, after this procedure we have seven maps (as listed in Table\,\ref{table:t2}) that allow us to relate the emission from the photosphere through the chromosphere and transition region to the surface magnetic field.
These are available at different spatial scales from 0.5 to 6\arcsec/pixel to test the role of spatial resolution, and for three different solar features, namely active region, quiet Sun, and coronal hole (cf.\,Table\,\ref{table:t1}).
In Fig.\,\ref{fig_a} we show some of these maps for the active region.

\subsection{Regions of interest}\label{S:roi}

To study the relation of the emission from the upper atmosphere to the magnetic field and the flux-flux relations we investigate different types of regions.
For these we will derive the correlations and power-law indices at different resolutions (see Sect.\,\ref{S:method}) and compare the properties of these different regions of interest.


\paragraph{(1)~ Active region (without sunspots or pores):}
This is almost the full field-of-view of the active region data set (cf.\ Fig.\,\ref{fig_a}). This region includes also relatively quiet areas and extended (enhanced) network areas, decaying active region (upper part of image) and a region of emerging flux (between the upper two yellow rectangles, associated with some of the strongest brightenings in \ion{Mg}{ii} k1r). However, sunspots and pores would significantly alter the general relations. For example in sunspots the \ion{Mg}{ii} lines differ significantly from the rest of the solar disc in that they do not show the self-reversal feature (h3 and k3), but they are singly peaked.
Therefore we define the active region as the area that encompasses the full raster map (red rectangle in Fig.\,\ref{fig_a}) excluding the regions covered by sunspots (yellow rectangles in Fig.\,\ref{fig_a}).
This is similar to the definition of an active region presented by \citet{1989ApJ...337..964S}.

\paragraph{(2)~ Plage: }
A large part of the full active region scan is covered by emission with very low intensity from the chromosphere and the transition region. Therefore, we exclude these data points for the definition of the plage regions.
For this we estimate the average and standard deviation, $\sigma$, in the images of the AIA\,1600\,\AA\ channel in a larger patch of a quiet region (orange rectangle in Fig.\,\ref{fig_a}).
We then define plage as the region where the AIA\,1600\,\AA\ emission is more than $2\,\sigma$ above the average; still excluding the sunspots. According to this definition, (enhanced) network areas and faculae are counted as plage.
This plage is shown by the green contours in Fig.\,\ref{fig_a}).

\paragraph{(3)~ Quiet Sun (QS):}
For comparison we also check the relations within quiet Sun regions.
Unfortunately we cannot use quiet regions from the active region data set shown in Fig.\,\ref{fig_a} for this analysis. This is because the exposure time is too short to give sufficient signal in particular in \ion{C}{ii} and \ion{Si}{iv}, and the quiet regions cover only a small portion of the field-of-view.
Therefore we located another large dense raster of IRIS that has sufficient signal-to-noise ratio (cf.\ Table\,\ref{table:t1}). Here the field of view is fully covered by quiet Sun with no active regions nearby.

\paragraph{(4)~ Coronal hole (CH):}
Just as for the quiet Sun we also compare the relations in a coronal hole region. Here again we have to investigate another data set and chose one where the IRIS raster was fully within a (on-disc) coronal hole (cf.\ Table\,\ref{table:t1}).


\paragraph{}
When investigating the relation of the magnetic field to the upper atmosphere emission, the mag-flux relations, we consider only those locations with an (absolute value of the) magnetic field strength of up to 200\,G.
This is the same threshold as used by \citet{2009A&A...497..273L}.
Other studies used higher thresholds, for example, \ \citet{1989ApJ...337..964S} used field strengths up to 800\,G.
However, all the four regions of interest considered here do not contain sunspots or pores, and those were also excluded in the study of \citet{1989ApJ...337..964S}.
Therefore there are only few data points left where the magnetic field strength as recorded by the moderately resolving HMI instrument is above 200\,G.
\citep[Of course, with a high-resolution instrument one can detect and resolve kilo-Gauss flux tubes even in the internetwork quiet Sun; e.g.][]{2010ApJ...723L.164L}.
To avoid the poor statistics for high magnetic field strengths and because the vast majority of the data points are to be found below 200\,G, we restrict the analysis to flux densities below this value.


\section{Methods}\label{S:method}

Based on the aligned images of the magnetograms, or rather of the absolute value of the pixel-averaged magnetic field strength $|B|$, and the intensities $I$ of the emission line features we study the relation between the emission from different parts of the solar atmosphere and the underlying magnetic field, here called mag-flux relations (Sect.\,\ref{S:power.laws})
For this we calculate a correlation coefficient and fit a power-law function to characterise the respective relations.
To get reliable mag-flux relations we first have to subtract the basal flux, that is, the emission that originates from the atmosphere in the absence of magnetic field (Sect.\,\ref{S:basal.flux}).
Finally we also study flux-flux relations, that is, the relation of the emission from different parts of the atmosphere (Sect.\,\ref{relations-ff}).

Later in Sects.\,\ref{S:results} and \ref{S:discussion} we will discuss different solar features over a range of spatial resolutions.
In the present section we describe the method using the example of the active region data set (Sect.\,\ref{S:roi}.1) at a spatial scale of 1.5\arcsec/pixel.

        \subsection{Basal flux}\label{S:basal.flux}
        
\begin{figure*}
   \sidecaption
   \includegraphics[width=12 cm]{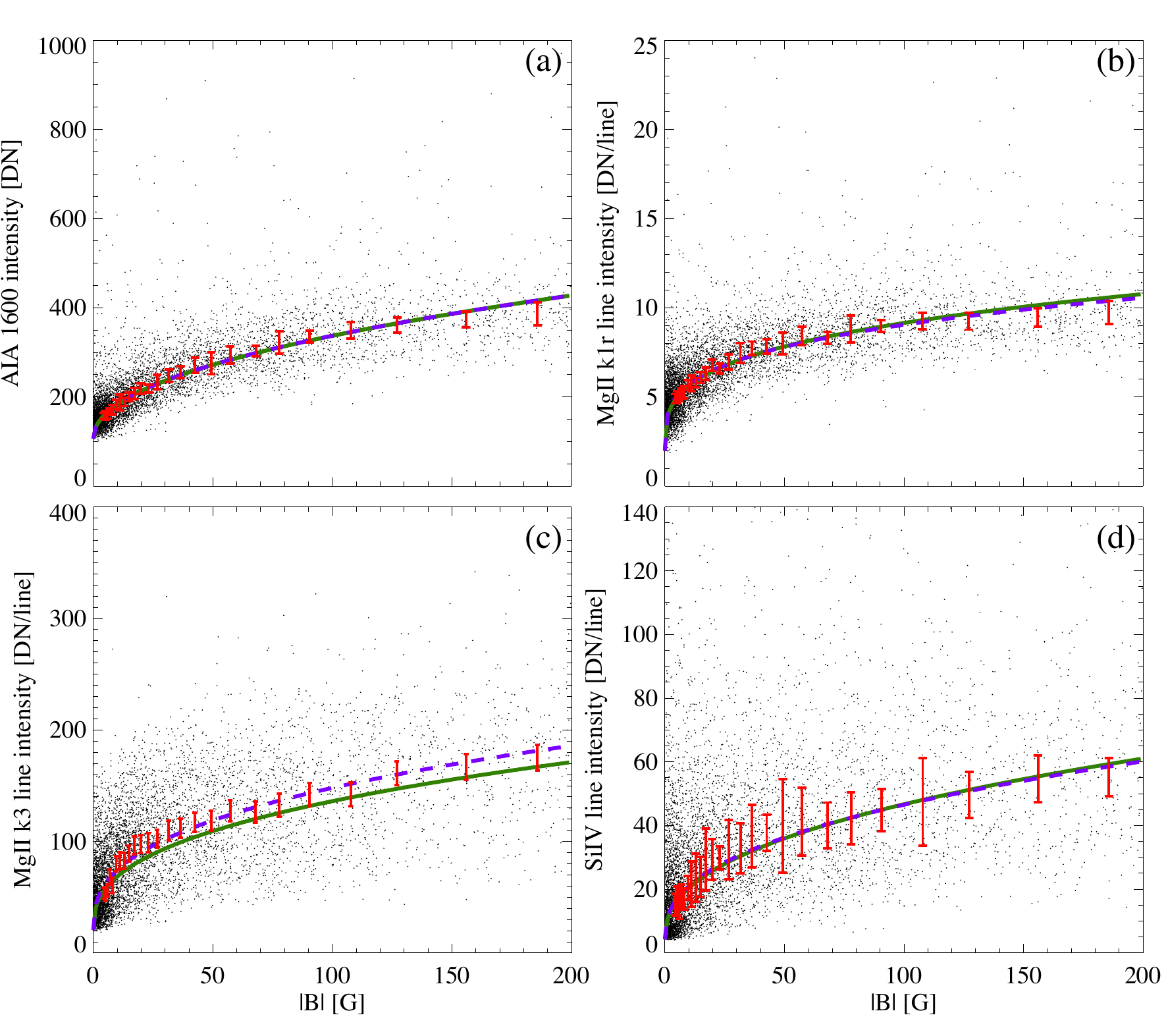} 
      \caption{Relation of upper atmosphere emission to the magnetic field.
While we prepared these plots to calculate correlations and power-law-fits for all emission features listed in Table\,\ref{table:t2}, for brevity we plot here only the relations involving (a) AIA 1600\,\AA, (b) \ion{Mg}{ii}\,k1r, (c) \ion{Mg}{ii}\,k3, and (d) \ion{Si}{iv}.
The power-law functions fitted to all data points are represented by (green) solid curves.
The red bars indicate the intensity in bins in magnetic field strength $|B|$, into which we grouped equal numbers of data points (here 200). The height of each bar represents three times the standard error in each bin. 
The blue dashed curves show power-law-fits to the binned data.
The data shown in these examples are for the active region set as defined in Sect.\,\ref{S:roi} (red rectangle without yellow rectangles in Fig.\,\ref{fig_a}).
The same data are plotted in Fig.\,\ref{F:BI.log.log} on a double-logarithmic scale. See Sects.\,\ref{S:method-cor} and \ref{S:power.law.fit}.}
              \label{fig_c}%
    \end{figure*}

       When determining the relation between the magnetic field strength and the intensity of a chromospheric or coronal emission line, for example, through a power law, the basal flux plays a critical role. This basal flux is the emission that is present even in the absence of magnetic field. Usually this is interpreted as representing emission from an atmosphere heated purely by acoustic waves and shocks \citep{1989ApJ...337..964S,1991SoPh..134...15R}.
In the scatter plots of radiative flux vs. magnetic flux density for a number of emission features in Fig.\,\ref{fig_c} it is clearly evident that there is a minimum radiative flux at low magnetic flux density.
Under the presence of magnetic field also the emission powered by magnetic heating of the atmosphere is present, and thus the radiative flux will increase with magnetic flux density.

The concept of basal flux was introduced by \citet{1987A&A...172..111S}  in the stellar context. Later, \citet{1989ApJ...337..964S} defined the basal flux simply as the minimum intensity in the field-of-view investigated on the Sun. Here we further evolve this concept. Implicitly, \citet{1989ApJ...337..964S} assumed that one would find the lowest intensity value only at the lowest magnetic field strengths which will be zero field strength somewhere in the region considered. Thus this would represent the basal flux. However, it is clear that this method is sensitive to outliers of the intensity maps. One can minimise this effect by investigating a time series of images \citep{1992A&A...258..507S}. However, this is not feasible for our data, which are maps produced by scanning with a spectrograph slit acquired over a comparably long time. Therefore, we modify the original procedure of \citet{1989ApJ...337..964S} to account for outliers.
       
        Our method to determine the basal flux is based on dividing the collection of pixels with low magnetic field strength into bins according to magnetic field strength. In each bin we then determine the minimum value of the intensity and define the median of these minimum values in the bins as the basal flux. In the absence of outliers this produces the same results as \citet{1989ApJ...337..964S}, but it can also cope with outliers. Of course, we repeat this procedure for each of the emission features (from AIA and IRIS; cf.\ Table\,\ref{table:t2}) to derive the basal flux separately for each line or spectral feature.
        
        Details of our method are given in Appendix\,\ref{S:bflux}. When calculating the basal flux we investigate only pixels in the images with magnetic field strengths below 4\,G, which is roughly half of the noise level of the HMI data we use for the magnetic field. This ensures that we only look at the regions that appear to be field-free (to HMI). It turns out that the best magnetic bin size for our analysis is 1\,G (see Appendix\,\ref{S:bflux}), which means that we subdivide the collection of low magnetic field pixels into four bins. The median of the minimum intensities in the bins is then used as the basal flux, $I_0$, when performing the power-law-fits in Sect.\,\ref{S:power.law.fit}.

  \subsection{Relating the upper atmosphere emission to the surface magnetic field: mag-flux relations}\label{S:power.laws}

        \subsubsection{Mag-flux: correlation} \label{S:method-cor}

With the basal flux subtracted we can now statistically relate the upper atmosphere emission to the magnetic field though the mag-flux relations.
In the resulting scatter plots of emission vs. magnetic field we find clear relations of the respective two quantities, in particular when comparing to the different features of the \ion{Mg}{ii} line and AIA\,1600\,\AA\ (see Fig.\,\ref{fig_c}).

Because these relations appear to be nonlinear we use the Spearman's rank order correlation coefficient (in short: Spearman correlation) for a quantitative analysis of this relation. The Spearman correlation measures any monotonic relation between two variables, regardless of the functional form, and is not sensitive to outliers.
This is in contrast to the cross-correlation coefficient that is frequently used, but strictly speaking works only to characterise a linear relation.
In general, for a nonlinear relation the cross-correlation will underestimate the correlation between the two quantities (for more details and examples see Sect.\,\ref{S:cor.flux.flux.result}).
Depending on the spatial resolution of the data the Spearman correlation coefficients are in the range from 0.6 to 0.9.
This shows that there is a clear relation that motivates the application of a power-law fit.

\subsubsection{Mag-flux: power-law relation}\label{S:power.law.fit}

The scatter plots (Fig.\,\ref{fig_c}) are clearly nonlinear, and plotting them on a double-logarithmic scale suggests a power-law trend between the magnetic field and intensities (cf. Fig.\,\ref{F:BI.log.log}). 
The same was noticed before for spatially resolved observations, for example, between \ion{Ca}{ii} and the magnetic field \citep{1989ApJ...337..964S} or 1600\,\AA\ images and magnetic field \citep{ 2009A&A...497..273L}.
Therefore we assume a power-law here, too, that is,
\begin{equation}  \label{eq1}
I'=I{-}I_{0}=a\cdot|$B$|^{b}~, 
\end{equation}
where $I$ is the intensity of the emission feature, $I_{0}$ the basal flux, $|B|$ the absolute value of the line-of-sight photospheric magnetic flux density, and $a$ a scaling parameter.
Most importantly, $b$ is the power-law-index.

Here we employ two methods of power-law-fitting to determine the power-law-index, $b$.
Three more methods are illustrated in Appendix\,\ref{S:pl-methods}, which all give similar results underlining the robustness of our findings.

In the first method we directly apply a least-squares fit to the power-law-function in Eq.\,\ref{eq1} to each of the scatter plots showing $I{-}I_0$ vs.\ $|B|$, where we apply statistical errors (Poisson weights) for the uncertainties in $I{-}I_0$.
The power law fits obtained this way are shown in Fig.\,\ref{fig_c} (green solid lines).

We also applied an alternative second method for the power law fitting in which we bin the data points in magnetic field strength.
For a consistent signal-to-noise we used bins with an equal number of data points (here 200) starting from 4\,G, that is, from above the level for the basal flux calculation (Sect.\,\ref{S:basal.flux}), to 200\,G.
The average intensities in these bins are indicated in Fig.\,\ref{fig_c} as bars, where the height of each bar represents (three times the)  standard error in the respective bin.
We apply the  power-law-fit to the average intensities in the bins with (the inverse of) the standard error in the bins as  weights. 
The resulting fits are shown in Fig.\,\ref{fig_c} (blue, dashed line) and they are virtually identical with the method of fitting all points directly (with the exception of \ion{Mg}{ii}\,k3, where nonetheless the slopes of the two fits are within a few percent).

The intensity values of the different line (features) show a scatter of up to a factor of about five in intensity when considering the regions with low magnetic field only (below 4\,G, see Fig.\,\ref{fig_c}). Part of this scatter is likely due to limitations of the spatial resolution. Should there be two patches of equally strong but opposite polarity magnetic field in one spatial resolution element of the magnetograph then, the resulting polarization and thus the derived magnetic flux density would be small. Still, in that area there would be considerable magnetic heating and thus a high chromospheric intensity. Consequently this pixel of the observation would be at low magnetic field, but high intensity. This underlines, that one has to use a measure of the minimum intensity at low magnetic fields to get a reliable measure of the basal flux. This determination of the basal flux through a minimum value, as originally suggested by \citet{1989ApJ...337..964S}, is based on the physical idea that the basal flux should represent the intensity in regions with almost no magnetic field. If one would use, for example, the average intensity at low magnetic fields, the result would be contaminated by small-scale mixed-polarity patches of magnetic field that cancel in the magnetograms.

\subsubsection{Basal flux and power-law relation}\label{bf_and_pow_law_rel}
To investigate the power-law relationship between magnetic field and intensity we first subtract the basal flux from the intensities, see Eq.\,\ref{eq1}. This has to be done because we want to study the relation of the magnetic field to the intensity, and therefore we have to exclude first the intensity from the field-free regions, which are heated by an independent process, likely dissipation of acoustic waves. In principle one could consider to do a fit between intensity and magnetic field without subtracting the basal flux first, but to have a constant offset as a free parameter of the fit. However, as outlined at the end of Sect.\,\ref{S:basal.flux}, the definition of the basal flux as a minimum intensity at low magnetic field strength is motivated by physics, and the offset in a fit would certainly not reproduce that value. Therefore we choose to first subtract the basal flux from the intensity. In this we follow not only the original work by \citet{1989ApJ...337..964S}, but also other studies such as by \citet{2007A&A...466.1131R}, \citet{1999ApJ...515..812H}, or \citet{2009A&A...497..273L, 2009IAUS..259..185L}.

\subsection{Flux-flux relations}\label{relations-ff}

The relations of the radiative fluxes from different parts of the atmosphere, the flux-flux-relations, contain valuable information on how the structure and possibly the governing processes change throughout the upper atmosphere.
For many stellar observations there is no information available on the magnetic field (distribution) on the stellar surface.
Even in those cases flux-flux relations are available as long as those stars are observed in different wavelengths originating from different parts of the upper atmosphere.

In principle, the relation between the magnetic field and the emission from the upper atmosphere, the mag-flux-relations, are directly linked to the processes governing the physics of the upper atmosphere.
However, when observing at high resolution, on the Sun the mag-flux relations might be obscured by the structure of the magnetic field.
The emission from the chromosphere and transition region originates from heights in the solar atmosphere of at least 1\,Mm and above.
For example the transition region is located over a height range from about 2\,Mm to 5\,Mm, intermittent both in time and space \citep[e.g.][]{2013SoPh..288..531P}.
Therefore even slightly inclined field originating from small magnetic flux tubes in the photosphere will prevent seeing a clear relation between magnetic field and emission when observing at a spatial resolution corresponding to 1\,Mm or better. 
Therefore we also investigate the flux-flux relation in the spatially resolved data.
In the following we will concentrate on the IRIS data alone, that is, on the emission from the features of \ion{Mg}{ii}, \ion{C}{ii} and \ion{Si}{iv}.
This has the advantage that these lines are observed through the same slit and the spatial alignment of the data is easily ensured through the fiducial marks on the slit \citep{2014SoPh..289.2733D}.

\begin{table}
\caption{Mutual relation of emission features for the active region set. The Spearman's Rank order correlation coefficients are presented in the blue part, the cross-correlation coefficients in the red part.\label{table:t3}}
\centering
\newcommand{\colR}{\cellcolor{red!15}}
\newcommand{\colG}{\cellcolor{green!15}}
\newcommand{\colB}{\cellcolor{blue!15}}
\begin{tabular}{cc@{~}|@{~}c||ccccc}
\hline\hline
&\multicolumn{2}{c||}{} & \multicolumn{3}{c}{\ion{Mg}{ii}} & & \tabularnewline
\cline{4-6}
&\multicolumn{2}{c||}{\raisebox{1ex}[-1ex]{correlation}} & k1r &  k2r &  k3 & \raisebox{1ex}[-1ex]{\ion{C}{ii}} & \raisebox{1ex}[-1ex]{\ion{Si}{iv}} \tabularnewline
\hline\hline
\hspace*{1em}
&                             & k1r&\colG1&\colB0.63&\colB0.54&\colB0.55&\colB0.48\tabularnewline
                &\ion{Mg}{ii} & k2r&\colR0.53&\colG1&\colB0.88&\colB0.84&\colB0.70\tabularnewline
&                             & k3 &\colR0.42&\colR0.88&\colG1&\colB0.81&\colB0.62\tabularnewline
&\multicolumn{2}{c||}{\ion{C}{ii}} &\colR0.24&\colR0.26&\colR0.22&\colG1&\colB0.80\tabularnewline
&\multicolumn{2}{c||}{\ion{Si}{iv}}&\colR0.22&\colR0.23&\colR0.17&\colR0.16&\colG1\tabularnewline
\hline\hline
\multicolumn{3}{c||}{power-law index}
&\_\_& 1.4 & 1.9 & 1.9 & 2.5
\tabularnewline
\multicolumn{3}{c||}{to \ion{Mg}{ii}\,k1r}
&& $\pm$0.2 & $\pm$0.01 & $\pm$0.2 & $\pm$0.4
\tabularnewline
\hline
\end{tabular}
\tablefoot{%
%
The correlation coefficients are based on the raster images at the original IRIS spatial resolution without the subtraction of the basal flux. See Sect.\,\ref{S:cor.flux.flux.result} for the correlation coefficients and Sect.\,\ref{S:method-flux} for the power-law indices. 
}
\end{table}

\subsubsection{Flux-flux: correlation}\label{S:cor.flux.flux.result}

Just as for the mag-flux relations, to quantify the relation between the different radiative fluxes from the upper atmosphere we use the Spearman correlation, because it is more appropriate for nonlinear relations.
The respective Spearman correlation coefficients for the active region set are listed in the upper right half of the correlation matrix in Table\,\ref{table:t3} (shaded blue).
As for the correlation to the magnetic field, here the correlation also drops with temperature.
For example in the top row of Table\,\ref{table:t3}, which shows the correlation to \ion{Mg}{ii}\,k1r, the correlation for \ion{Mg}{ii}\,k2r is higher than the one for \ion{Si}{iv}.

To illustrate the difference between the Spearman correlation and the cross-correlation, in Table\,\ref{table:t3} we also give the respective cross-correlation coefficients (lower left part of the matrix, shaded red).
As mentioned in Sect.\,\ref{S:method-cor} in general the cross-correlation coefficients are significantly lower than the Spearman correlation, which is basically because the relations between these quantities are nonlinear.
We note that in their study \citet{2015ApJ...811..127S} found a correlation between \ion{Mg}{ii}\,h1v to h2v of only 0.33 for the quiet Sun using a cross correlation technique which is lower than the 0.38 we find using the Spearman correlation for the quiet Sun set.
This underlines the importance of using the Spearman correlation in the presence of nonlinear relations.

\subsubsection{Flux-flux: power-law relation}\label{S:method-flux}

\begin{figure}[h!]
   \centering
   \includegraphics[width=8.8 cm]{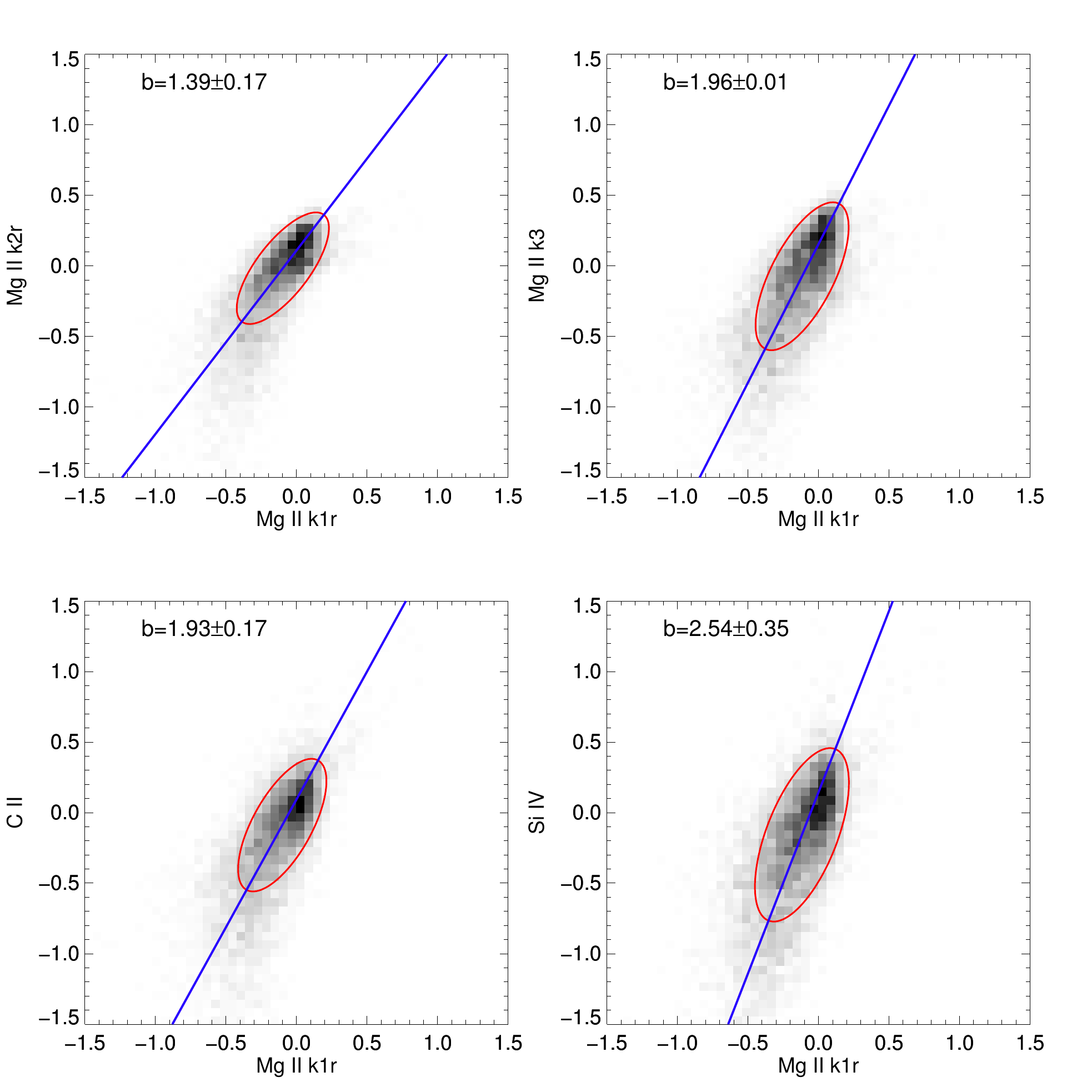}
   \caption{Probability density functions (PDFs) for the flux-flux relation
   for the active region set  for a spatial resolution of 1.5\arcsec. Here the respective basal fluxes are subtracted. The intensities are normalised to the respective median value and are plotted on a logarithmic scale. The red ellipses show the fit to the contour level at 1.5$\,{\times}\,$FWHM. The blue lines indicate the power-law-slopes calculated as the average from the ellipse fits of the contour levels at [1,1.5,2]$\,{\times}\,$FWHM. See Sect.\,\ref{S:method-flux} 
   }
              \label{fig_e}
    \end{figure}

To derive the power-law-index for the flux-flux relations we take another route than for the mag-flux relations, where we fit directly power-law functions to the observables.
When performing the power-law fits for the flux-flux relations we encountered several problems for lines that show a large scatter in intensity. In those cases, the power-law relations derived by the direct fitting procedure sometimes did produce results that did not look consistent with the data, mainly caused by outliers.
Here a closer look at the scatter plots of the flux-flux relation, or more precisely the probability density functions (PDF), indicates a way to a more robust technique: The PDFs show iso-contour levels that are close to ellipses (see Fig.\,\ref{fig_e}).
Clearly, for two well-correlated quantities the semi-major axis of the ellipse fitting the PDF will provide the slope of the relation.
%
%
Here, the basal flux is subtracted first, of course. We fit an ellipse to the PDFs (on log-log scale), with the major axis then representing the slope, that is, the power-law index.
Thus, when deriving the flux-flux relations of the different regions for the different spatial resolutions we used the power-law indices derived from the ellipse fitting.

More importantly, the direct fitting of a power-law (as in Sect.\,\ref{S:power.law.fit}) gives results very similar to the ellipse fitting (cf. Appendix \ref{S:pl-methods}), as long as there is not too much noise and the power-law-fits do not fail. This is demonstrated for the mag-flux relations (method V in Appendix\,\ref{S:pl-methods}).
While we do not give a full theoretical justification for the ellipse fitting method, the comparison to other methods shows that the ellipse fitting gives reliable results.

To obtain a power-law-index we use a two-step procedure. First we fit ellipses to 50 contour levels of the PDF ranging from 25\% to 75\% of the peak value of the PDF. We then fit a Gaussian along the average major axis of the fitted ellipses to determine the full width at half maximum (FWHM) of the PDF. Finally we fit three ellipses to the PDF at the contour levels with $[1,1.5,2]\,{\times}\,$FWHM. The average of the major axis for these three ellipses defines the slope of the power-law function, the difference between the minimum and maximum values we use as an estimate for the error of the slope.

The results for this procedure in the case of the active region set are illustrated in Fig.\,\ref{fig_e}.
Here we plot the fitted ellipses (to 1.5$\,{\times}\,$FWHM) for each of the flux-flux relations along with the power-law (see also Table\,\ref{table:t3}).
This same procedure is applied  to all the different regions-of-interests for the different spatial resolutions, just as for the mag-flux relations.

\section{Results} \label{S:results} 

Here we first discuss the results for the active region set before we compare these to the plage, quiet Sun and coronal hole sets. For the definition of these regions of interest see Sect.\,\ref{S:roi}.

   \subsection{Active region (without sunspots and pores)}\label{S:pl.results} 

\subsubsection{The mag-flux relations}\label{S:cor.pow.plage.results} 

        The majority of previous studies of the (spatial) relation between magnetic field and intensity were based on lower spatial resolution data than used in this paper. However, \citet{2011A&A...531A.112K} and \citet{2017ApJS..229...12K} used better quality magnetograms than our study. To compare our results with previous results and to study the impact of spatial resolution on correlation coefficients and power-law-indices we investigate the data sets at  spatial samplings from 0.5\arcsec\ to 6.0\arcsec\ per pixel. For each resolution the basal flux was calculated separately and then subtracted from the intensities to derive the correlation and power-law indices.
       
\begin{figure*}
\begin{center}
\resizebox{0.495\hsize}{!}{\includegraphics{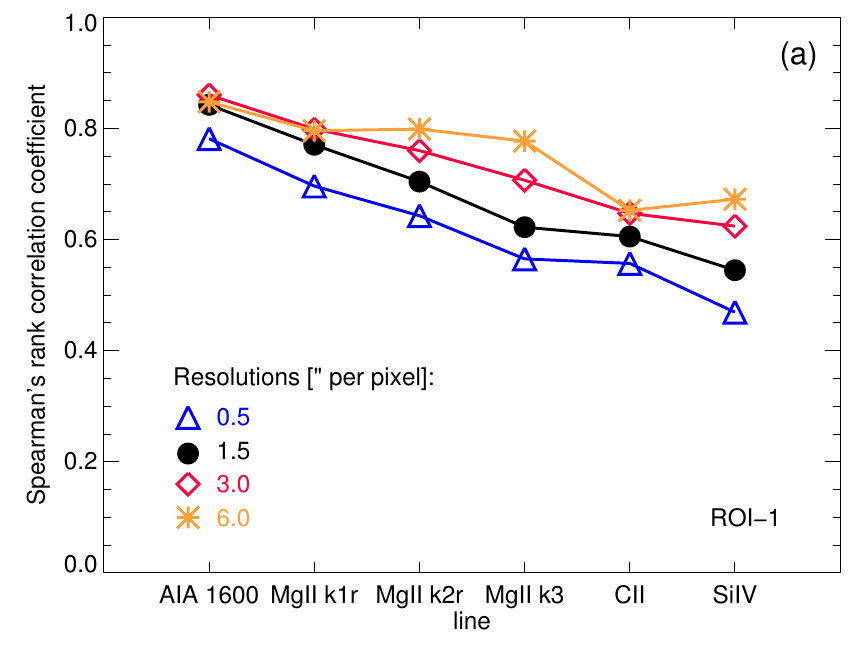}}
\resizebox{0.495\hsize}{!}{\includegraphics{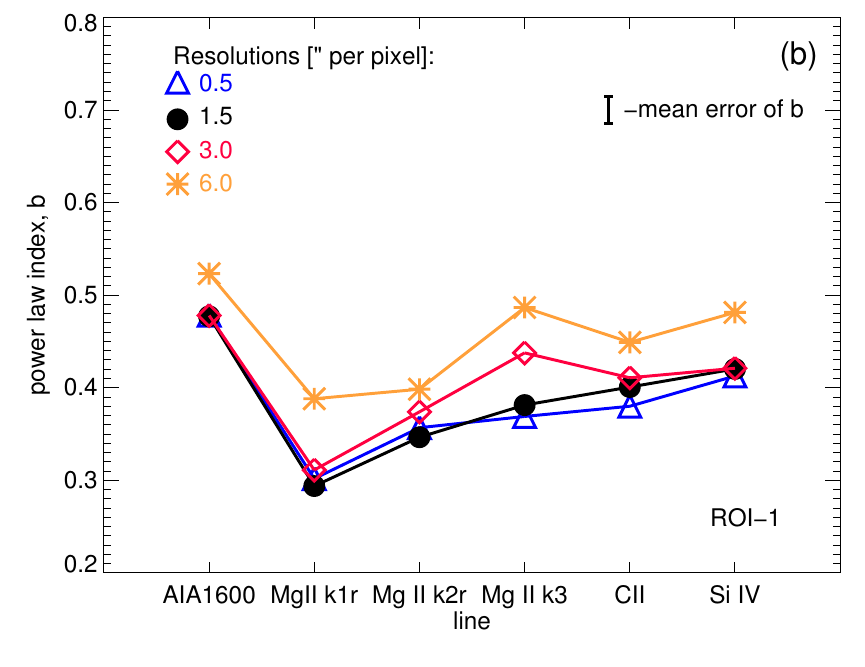}} 
\caption{Relation of the upper atmosphere to the underlying magnetic field for the active region set. Panel (a) shows the Spearman correlation of the various emission features to the photospheric magnetic field, panel (b) the power-law index from the power-law fit of the relation of emission to magnetic field. The emission features are ordered according to the approximate formation temperature (cf.\ Table\,\ref{table:t2}).  The different colours show the results based on data at different spatial resolutions, here represented by plate scales from 0.5\arcsec\ to 6\arcsec\ per pixel. See Sect.\,\ref{S:cor.pow.plage.results}. \label{fig_f}} 
\end{center}
\end{figure*}

       The correlation between the emission and the underlying magnetic field (cf.\ Sect.\ref{S:method-cor}) drops monotonically with temperature (in an average sense; see Table\,\ref{table:t2} for the temperatures). This trend is illustrated in  Fig.\,\ref{fig_f}a that shows the Spearman correlation between emission and magnetic field for four spatial resolutions. This trend is seamlessly connecting also across instruments from the AIA\,1600\,\AA\ channel to \ion{Mg}{II}\,k1r observed with IRIS.  When checking different spatial resolutions, we find the trend of the correlation from the temperature minimum to the transition region to be independent of resolution.
However, for the same emission feature we see a higher correlation at lower spatial resolution.  This latter result is consistent with the \citet{1989ApJ...337..964S} analysis that concentrated on the \ion{Ca}{ii} line.
We will discuss a physical scenario for this drop of the correlation coefficient in Sect.\,\ref{S:correlation.discussion}.

The power-law index characterising the relation between the emission features and the magnetic field (cf.\ Sect.\,\ref{S:power.law.fit}) shows a more peculiar behaviour. The index grows monotonically from \ion{Mg}{II} to the transition region and will be discussed in Sect.\,\ref{S:power.law.discussion}.
However, the power-law index for the AIA\,1600\,\AA\ channel is higher than for the other emission features, giving the plot in Fig.\,\ref{fig_f}b the appearance of a hockey stick.
In Sect.\,\ref{S:aia1600.results} we will show that the drop from AIA\,1600\,\AA\ to \ion{Mg}{ii}\,k1r is real and actually
can be smoothly followed through the temperature minimum region. This is why we show connecting
lines between the data points for AIA\,1600\,\AA\ and \ion{Mg}{ii}\,k1r in Fig.\,\ref{fig_f}b.

 This same hockey-stick trend of the power-law index is present irrespective of spatial resolution. For resolutions from 0.5\arcsec\ to 3\arcsec\ per pixel we find almost the same power-law indices. This is in agreement with \citet{1989ApJ...337..964S}
who found the power-law index to be independent of resolution (in the case of \ion{Ca}{II} H and K). However, for coarse resolutions we find a slight increase of the power-law index for all the emission features that seems to be larger than the error bars (Fig.\,\ref{fig_f}b).

       While the correlation and (to a lesser degree) the power-law index depend on the spatial resolution, the overall trend remains unaffected by the resolution. 
Therefore the variation of these parameters throughout the atmosphere from the temperature minimum through the chromosphere into the transition region is a robust result. This includes the monotonic drop of the correlation coefficient as well as the hockey-stick-type trend of the power-law index.

       \subsubsection{The flux-flux relations}\label{S:flux.flux.results} 

      In order to be independent of the impact that combining different instruments and possible misalignments might have, we also study the flux-flux relations of the emission features seen by the IRIS spectrograph alone (cf.\ Sect.\,\ref{S:method-flux}). When looking at the power-law indices of the  emission lines (or their features) with respect to \ion{Mg}{ii} k1r (bottom row of Table\,\ref{table:t3})
we find an increase with temperature. This confirms the increase of the power-law index for the mag-flux relations for spectral lines forming at temperatures higher than the equivalent of \ion{Mg}{ii} k1r
(i.e. the long handle of the hockey stick).

   \begin{figure}[h!]
   \centering
   \includegraphics[width=8.8 cm]{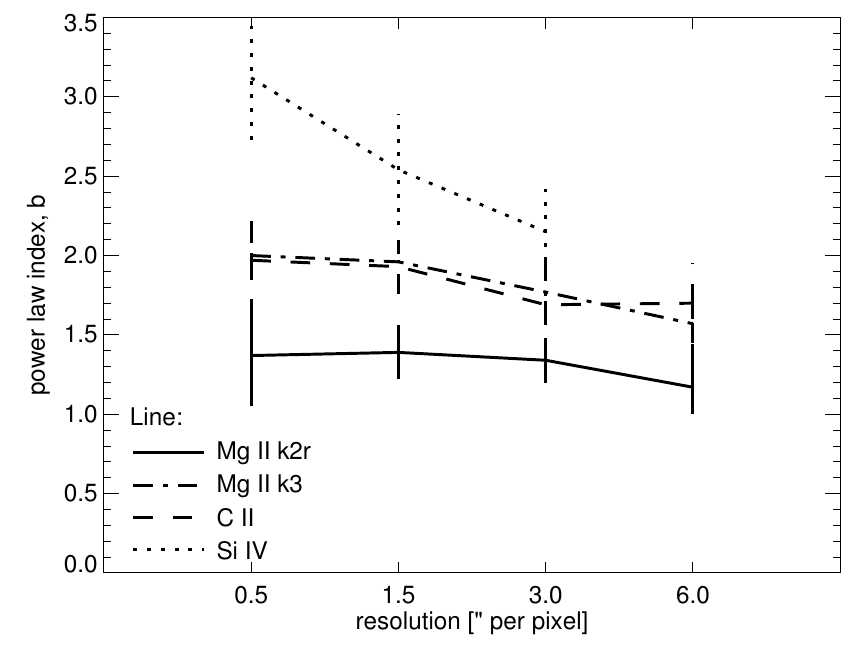} 
   \caption{Power-law index of the flux-flux relations for different resolutions (here as plate scale). All the power-law indices are derived from the ellipse fits to the double-logarithmic plots (cf.\ Fig.\,\ref{fig_e}) of the respective emission line (feature) versus \ion{Mg}{ii}\,k1r for the active region set. See Sect.\,\ref{S:flux.flux.results}.}  
              \label{fig_g}%
    \end{figure}

When handling the IRIS data alone, the spatial alignment of the emission features can be done with very high precision, which is ideal for a comparison on how the relations change with spatial resolution. When comparing different instruments a difference between fine and coarse resolution could also originate in imperfections of the alignment --- which is eliminated here by using IRIS-only data. In Fig.\,\ref{fig_g} we show how the power-law indices of the emission features with respect to \ion{Mg}{ii} k1r
change with spatial resolution. For the other \ion{Mg}{ii} features as well as for \ion{C}{ii} there is no change with spatial resolution. However, for \ion{Si}{iv} a clear trend can be seen with a significantly steeper power-law index at
high spatial resolution.
(Even though the results for \ion{Si}{iv} were less reliable than for the other lines because of the small number of data points when binning).
The possible implications of this will be discussed in Sect.\,\ref{S:flux.flux.discussion}.

We also checked if the different emission features are independent quantities by comparing the power-law indices of the mutual power law relations.
If they are independent, then the following should apply to the intensities of the different lines:
\begin{equation}
\left.
    \begin{array}{r@{~}c@{~}l}
    \mbox{C\,\sc{ii}} &\propto&  \mbox{Mg\,\sc{ii}}^{\,\alpha}
    \\[0.5ex]
    \mbox{Si\,\sc{iv}} &\propto&  \mbox{Mg\,\sc{ii}}^{\,\beta}
    \end{array}
\right\}
\quad
	 \longrightarrow
\quad
\left\{
    \begin{array}{l}
    \mbox{Si\,\sc{iv}} \propto  \mbox{C\,\sc{ii}}^{\,\gamma}
    \\[0.5ex]
    ~~~~~~~~\mbox{with}\quad \gamma=\,\beta/\alpha.
    \end{array}
\right.
\end{equation}
For the active region set (1.5\arcsec\ plate scale) from the values of $\alpha$ and $\beta$ in Table\,\ref{table:t3} (and Fig.\,\ref{fig_e}) we find $\beta/\alpha{=}1.3 $. This is close to the value of $\gamma=1.21{\pm}0.12$ derived directly from the flux-flux relation between \ion{Si}{iv} and \ion{C}{ii} (and within the error estimate). Also for other regions of interest and spatial resolutions the values for $\beta/\alpha$ and for $\gamma$ differ by less than the error estimate.

This result implies that the scatter of the intensity in the flux-flux relations is independent of the intensity itself.\footnote{The level of the photon noise is significantly smaller than scatter which we observed.} 
In  Sect.\,\ref{S:flux.flux.discussion} we discuss what this implies for the upper atmosphere.

    \subsection{The plage area excluding the quiet Sun}\label{S:only.plage.results} 

    \begin{figure*}
\begin{center}
\resizebox{0.495\hsize}{!}{\includegraphics{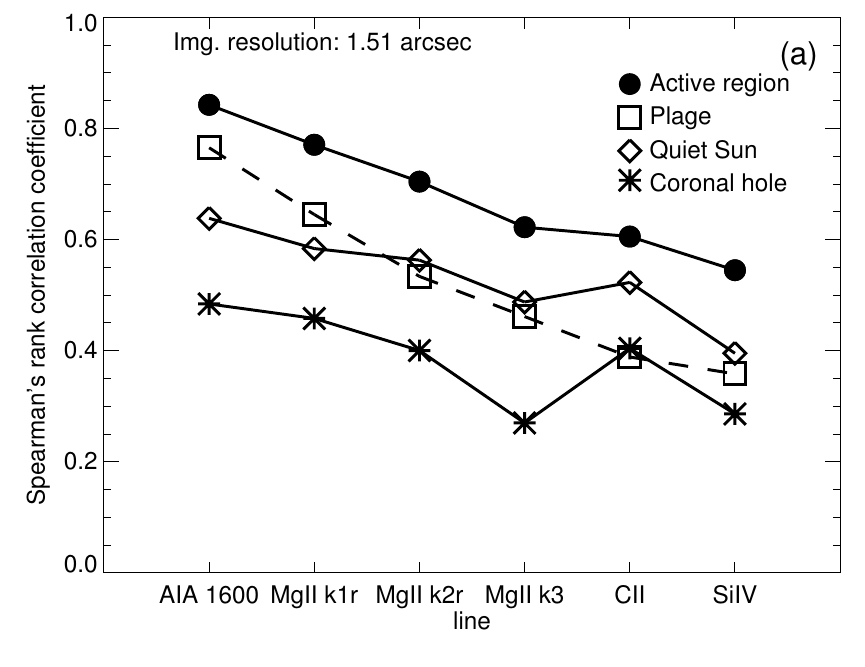}} 
\resizebox{0.495\hsize}{!}{\includegraphics{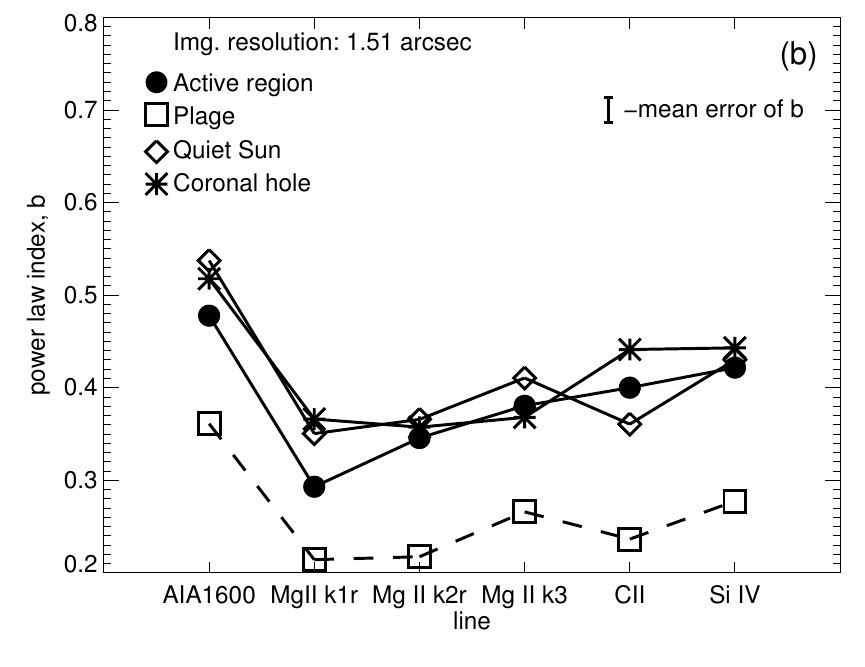}} 
\caption{Relation of the upper atmosphere to the underlying magnetic field in different parts of the Sun. This figure is similar to Fig.\,\ref{fig_f}, but  for a spatial scale of 1.5\arcsec\ per pixel for the four regions of interest as defined in Sect.\,\ref{S:roi}. Panel (a) shows the Spearman correlation to  the (underlying) magnetic field and panel (b) the power-law index of the power law fit of emission feature vs. magnetic field. The curves for active region (solid with filled dots) are identical to the respective black curves in Fig.\,\ref{fig_f}. See Sects.\,\ref{S:only.plage.results} to \ref{S:qs.results}.
\label{fig_h}} 
\end{center}
\end{figure*}

 In general the results for the plage area as well as those for the quiet Sun and the coronal hole are similar to the active region. Therefore in this and the following subsection we concentrate mainly on how the results in these other parts of the Sun differ from the active region.

The plage region is defined by excluding the more quiet parts from the active region domain (cf.\ Sect.\,\ref{S:roi}). Basically this leaves the parts of the active region that are outside sunspots and pores where the chromosphere appears bright.

 There are two major differences to the active region: (1) The correlation of the emission features to the magnetic field seems to drop faster with formation temperature. (2) While
the overall hockey-stick-type trend of the power-law-index remains unchanged, the power-law indices are significantly lower than in the active region set.
These findings are discussed in Sects.\,\ref{S:correlation.discussion} and in Sect.\,\ref{S:power.law.discussion}.

    \subsection{Quiet Sun and coronal hole}\label{S:qs.results} 

The
results for the quiet Sun and the coronal hole differ mainly in two aspects from the active region data set. The correlations are lower while the power-law indices are still comparable to the active region (without sunspots and plages).

For the correlation between the emission and the magnetic field we see a clear ordering, with the Spearman correlation being lower in quiet sun compared to the active region (and comparable to plage), while the correlation in the coronal hole is even less (Fig.\,\ref{fig_h}a).
However, the trend of the correlation with temperature is similar in all cases in that the correlation drops with formation temperatures. This will be discussed in Sect.\,\ref{S:correlation.discussion}.

In contrast to the correlation coefficients, the power-law indices for quiet Sun and coronal holes are comparable
to those in the active region, with respect to their values for each emission feature as well as for the trend with formation temperature (Fig.\,\ref{fig_h}b).
This is surprising, because for the plage regions the lower correlation coefficients go along with reduced power-law indices (cf.\ Sect.\,\ref{S:only.plage.results}), as will be discussed in Sect.\,\ref{S:power.law.discussion}.

\section{Discussion}\label{S:discussion}
The main goal of this study is to shed light on the connection between different temperature regimes in the upper solar atmosphere and the underlying magnetic field. For this purpose, we discuss correlations (Sect.\,\ref{S:correlation.discussion}) and power-law relations (Sects.\,\ref{S:aia1600.results} to \ref{S:mag.flux.diff.regions}) 
between magnetic field and intensities as well as the mutual relation between intensities (Sect.\,\ref{S:flux.flux.discussion}).

Studies of the relation of the surface magnetic field to the emission  have been conducted before, for example, to \ion{Ca}{ii} \citep{1989ApJ...337..964S} or the UV continuum \citep{2009A&A...497..273L}.
However, these studies did not investigate how the relation of the magnetic field to the emission is changing throughout the chromosphere and into the transition region, an aspect that we concentrate on in this work.

\subsection{Correlation between magnetic field and emission}\label{S:correlation.discussion}

Our observations show that the  correlation with the underlying magnetic field decreases with increasing formation temperatures of the emission feature through the upper atmosphere (Sect.\,\ref{S:cor.pow.plage.results}, Fig.\,\ref{fig_f}a).     The new aspect we highlight here is that a continuous drop of the correlation with the magnetic field is found even within the chromosphere over the formation regions of the \ion{Mg}{ii} line features, from k1 to k2 to k3 --- in a statistical sense. The high spatial complexity and the intermittent nature of the chromosphere will cause  significant deviations from the average correlation, which is clear from the large scatter that is much larger than the measurement errors (seen when plotting, e.g. the emission vs.\ the magnetic field as in Fig.\,\ref{fig_c}).  

 The simplest explanation of the drop of the correlation is based on the magnetic expansion of the chromosphere \citep[e.g.][]{1990A&A...234..519S} that will also be instrumental to interpret the power-law relations later in Sect.\,\ref{S:power.law.discussion}. Starting from small patches in the photosphere, the magnetic flux tubes expand and cover an increasing fraction of the hot horizontal plane towards higher altitudes, which in a 1D semi-empirical model  \cite[e.g.][]{1981ApJS...45..635V} translates  to layers of higher temperature. Therefore, the emission in the chromosphere forming at increasingly higher temperatures will fill a larger and larger fraction of the surface area. Naturally, then the correlation between the magnetic field and the emission will decrease with increasing temperature. In addition, the flux tubes can (and many will) be inclined, the inclination often increasing with height (if the field is part of a loop), so that there will be a spatial mismatch between magnetic field and emission. This effect increases with formation height of the emission \citep[in space and temperature if considering a simple stratification
like in][]{1981ApJS...45..635V}.

For the emission forming at temperatures higher than in the chromosphere, that is, the transition region, we can expect even less correlation.  For lines such as \ion{C}{ii} or \ion{Si}{iv} at least part of the emission will form at the footpoints of loops reaching higher up \cite[see e.g.][]{2001A&A...374.1108P}. As most of these loops will be inclined, this emission will be, in general, mismatched to the underlying magnetic field.
Therefore, the decreasing correlation from the low chromosphere into the transition region can be well understood by the geometry of the flux tubes hosting the emitting plasma. Still, it is surprising that despite of all the spatial and temporal variability this average trend of a dropping correlation is so clearly defined.

The drop of the correlation with increasing spatial resolution (Fig.\,\ref{fig_f}a) can be also understood by geometric effects, again in part based on the flux-tube picture. Firstly, when the spatial resolution decreases, the small footpoints in the photosphere get smoothed out, which leads to a better correlation to the upper atmosphere, where the emission forms larger patches because of the fanning out of the flux tubes. Secondly, at lower resolution also flux tubes with an  inclination to the line-of-sight will show a match between their footpoints and the emission higher up, leading to higher correlation. And thirdly, when observing at lower resolution, many of the small-scale opposite polarity magnetic concentrations will cancel in the polarisation signal resulting in a smoother magnetic field map, which in turn gives a better correlation to the emission in the fanned-out part of the flux tubes.

In their original work, \citet{1989ApJ...337..964S} suggested that a typical spatial scale should exist because they found the scatter about the mean relation of (\ion{Ca}{ii}) emission vs.\ magnetic field to be independent of spatial resolution. The scatter only dropped once they degraded the spatial resolution to about 14\arcsec. However, it is not clear how such a typical length scale for the magnetic field patches should be compatible with our finding that there is a continuous drop of the correlation between emission and magnetic field from sub-arcsec resolution to several arcsec. \citet{1989ApJ...337..964S} also suggested that the increase of the correlation when degrading the resolution could be understood by the resolution element becoming comparable in size to the characteristic length of the chromospheric oscillation. However, here we wonder how this effect should explain the decrease of the correlation also for the transition region lines, which are not heavily affected by oscillations seen in the chromosphere.

At first sight it is surprising that in general the correlation for the plage area is slightly lower than for the active region and also drops faster with formation temperature (Fig.\,\ref{fig_h}a). However, in our definition  the plage region is the part of the active region (without sunspots and pores) that excludes the parts where the chromosphere is dark (see Fig.\,\ref{fig_a}, Sect.\,\ref{S:roi}), but includes emerging regions and an arch filament system.
Leaving out the darker part of the chromosphere makes the emission maps more homogeneous and thus the contrast to the highly structured magnetic field becomes less. Consequently the correlation coefficients will be lower.

In the quiet Sun and the coronal hole data sets the correlation is systematically lower than in the active region (Fig.\,\ref{fig_h}a), too.
This is to be expected, because in the quiet Sun and coronal hole there will be a smaller fraction of magnetic structures reaching up into the chromosphere and transition region. Thus unlike in the active region, there is a less clear connection from the photosphere to the upper atmosphere. The smaller coverage by magnetic features in the photosphere allows these to expand  more strongly in the chromosphere before they are stopped by a neighbouring magnetic feature. The quiet Sun photosphere and chromosphere is filled with low-lying loops that also reduce the correlation with the underlying magnetic field \citep{2010ApJ...723L.185W, 2013SoPh..283..253W}.

 Interestingly, the correlation in the coronal hole is significantly lower than for the quiet Sun, even low in the chromosphere. This is a surprise because in general the differences between coronal hole and quiet Sun are considered to be small at chromospheric levels. However, \citet{2004SoPh..225..227W} showed through a potential field extrapolation that in a coronal hole magnetic fieldlines typically reach less high and are shorter. This could imply a weaker relation of the surface magnetic field to the chromosphere and transition region  because more fieldlines close back to the Sun before they reach the chromosphere. Although this might explain the small correlation we see in a coronal hole, clearly further work is needed to better investigate the (magnetic) differences of coronal hole and quiet Sun in the chromosphere and the transition region. The active region considered in this study is young and emerging. It would be interesting to study how the power-law behaviour changes as a function of active region age.

\subsection{Mag-flux relation from the photosphere to the chromosphere}\label{S:aia1600.results}

\begin{figure*}
\begin{center}
\resizebox{0.495\hsize}{!}{\includegraphics{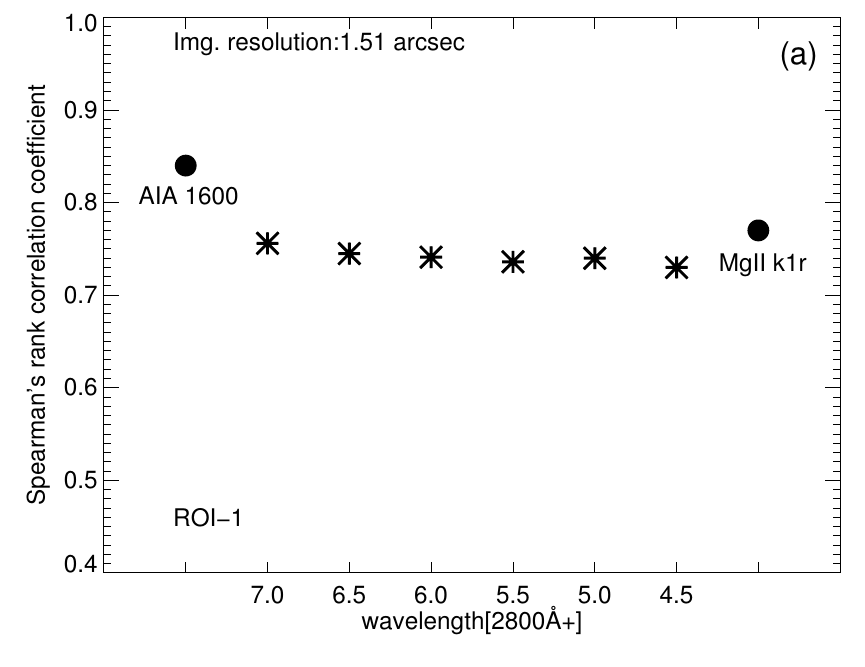}}
\resizebox{0.495\hsize}{!}{\includegraphics{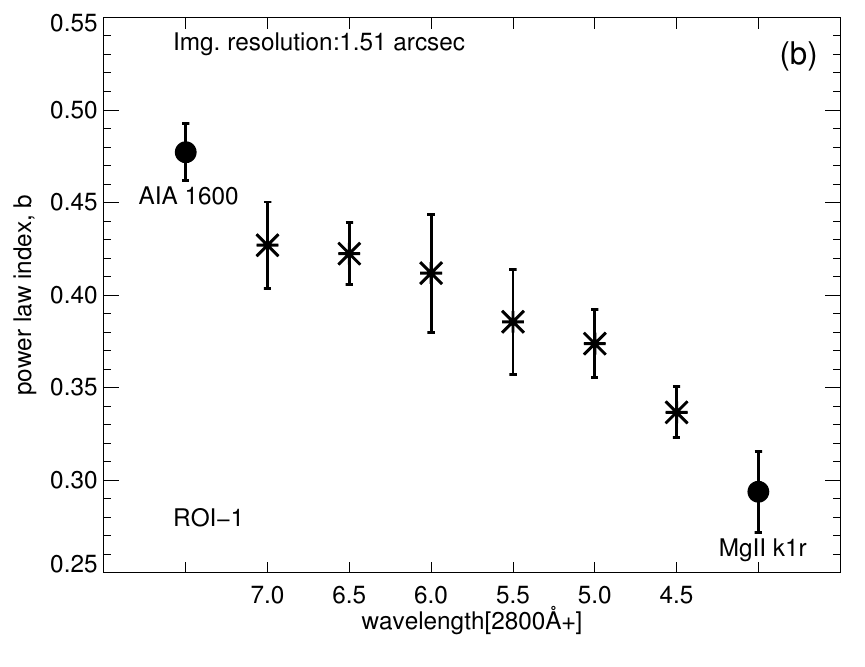}}
\caption{Variation of correlation coefficients and power-law indices from photosphere to temperature minimum. Using the \ion{Mg}{ii} wing this plot fills the gap between the formation region of AIA\,1600\,\AA\ and \ion{Mg}{ii}\,k1r in Figs.\,\ref{fig_f} and
\ref{fig_h}.  Panel
(a) shows the Spearman correlation and panel (b) the power-law index of the
relation of the emission to the magnetic field. The asterisks show results based on the emission in 0.5\,\AA\ wide bands in the red wing of \ion{Mg}{ii}\,h. The formation height increases with decreasing wavelength (i.e. to the right). These data are based on the active region set with a spatial scale of 1.5\arcsec, thus the black filled circles show the same data points as the black filled circles for AIA\,1600\,\AA\ and \ion{Mg}{ii}\,k1r in Figs.\,\ref{fig_f} and \ref{fig_h}. See Sect.\,\ref{S:aia1600.results}. 
\label{fig_s}} 
\end{center}
\end{figure*}
     
The overall variation of the power-law index of the relation of magnetic field to the emission shows a characteristic hockey-stick shape when plotted with respect to the formation temperature. This effect is present, irrespective of spatial resolution (Fig.\,\ref{fig_f}) or region of interest (Fig.\,\ref{fig_h}). Before discussing the physical cause of the sharp drop from the AIA\,1600\,\AA\ channel to the \ion{Mg}{ii}\,k1r feature we clarify that this drop is real and not just an artefact of the observation
\footnote{One might also speculate that the power-law index of the AIA\,1600\,\AA\ channel is higher than for \ion{Mg}{ii} because of the contribution of \ion{C}{iv} to the 1600\,\AA\ channel. However this is not the case. Firstly, the power-law index for \ion{Si}{iv} (that should be comparable to \ion{C}{iv)} is lower than for the 1600\,\AA\ channel. Secondly, we find almost the same results for the 1700\,\AA\ channel that has much less (probably close to none) contamination by \ion{C}{iv}.}, for example, caused by combining different instruments.

In order to investigate the change from AIA\,1600\,\AA\
 to  \ion{Mg}{ii}\,k1r
we study the wing of the \ion{Mg}{ii} line. According to \cite{1981ApJS...45..635V} k1r forms at the temperature minimum and further out in the wing the emission originates from lower heights in the photosphere (assuming a static atmosphere). Here we investigate the \ion{Mg}{ii} wing from 2804.5\,\AA\ to 2807.0\,\AA\ in 0.5\,\AA\ steps, each integrated over a 0.5\,\AA\ wide pass band.%
\footnote{For convenience we look at the red wing of \ion{Mg}{ii}\,h here instead of the k-line used before. The h1r minimum is found at about
2804\,\AA.  We use the h-line wing simply because of the location of the line on the detector within the wavelength window transmitted from the spacecraft: in the case of the red wing of the h-line we can go further into the continuum. The results do not change if we use (a more limited range) with the k-line.}
This provides information on the upper photosphere with longer wavelengths corresponding to lower heights. We treat the raster maps in these narrow bands in the same way as we treated the other emission features (cf.\ Sect.\ref{S:power.laws})
and derive the Spearman correlation and the power-law index of the relation to the surface magnetic field. The results are shown in Fig.\,\ref{fig_s}. Assuming that the AIA\,1600\,\AA\ channel forms below the \ion{Mg}{ii} wing (see  below) the abscissa in Fig.\,\ref{fig_s} is ordered by formation height (increasing to the right). While Fig.\,\ref{fig_s} shows the results for the active region set (at 1.5\arcsec\ spatial scale) the results for the other data sets (and resolutions) are qualitatively the same.

The correlation and power-law indices in the \ion{Mg}{ii} wing fill the gap between AIA\,1600\,\AA\ and \ion{Mg}{ii}\,k1r
quite well. The Spearman correlations of AIA\,1600\,\AA\ and \ion{Mg}{ii}\,k1r to the magnetic field are quite similar (cf.\ Figs.\,\ref{fig_f}a,\,\ref{fig_h}a), and so are the correlation coefficients in the wing of \ion{Mg}{ii} (Fig.\,\ref{fig_s}a).
The main interest is in the variation of the power-law index, and in particular concerning the question if there is a smooth transition on the short end of the hockey stick between AIA\,1600\,\AA\ and \ion{Mg}{ii}\,k1r (cf.\ Figs.\,\ref{fig_f}b,\,\ref{fig_h}b). Here the wing of \ion{Mg}{ii} provides a continuous connection emphasising that the power-law index drops smoothly from the upper photosphere to the temperature minimum (Fig.\,\ref{fig_s}b). This result underlines that the form of the hockey-stick variation is not just a measurement error or an effect of combining different instruments.

In the following subsections we discuss four possible scenarios for the drop of the power-law index with temperature from the (upper) photosphere to the temperature minimum at the base of the chromosphere. The four effects that can contribute to the observed behaviour are related to (i) the dependence of the merging height of flux tubes on the magnetic flux density; (ii) the height of formation of the radiation; (iii) the wavelength dependence of the intensity-temperature relationship, and (iv) the dependence of flux-tube heating on magnetic flux density.

\subsubsection{Geometrical model of magnetic flux
tube expansion and merging}\label{S:pli.flux.merging}

As mentioned in the introduction, originally \cite{1989ApJ...337..964S}
suggested that the geometric expansion of magnetic
flux tubes could explain the nonlinear power-law variation
of the intensity with increasing magnetic field with
a power-law index below one (see their Sect.\,VII.c).
However, they did not discuss if or how the wine-glass
shape of the magnetic field would cause the power-law
index to change with temperature; they were only concerned
with \ion{Ca}{ii}. Still one might wonder if an extension
of that model could provide some explanation for the
trend of the power-law index we see below the temperature
minimum.

We briefly recall the original idea from \cite{1989ApJ...337..964S}.
They suggested that where
the (average) magnetic field is low, neighbouring
flux tubes merge at  comparably high altitudes, and
the  chromospheric emission originates from below that
height of merging.
Packing more flux tubes into the same area (i.e.\ increasing
the average field) first would lead to a linear increase
of the emission with the magnetic field. Once the packing
gets too dense, the expanding flux tubes start merging below
the height where the chromospheric emission originates,
so further increasing the packing will not (significantly)
increase the emission intensity from the chromosphere:
The intensity will be related to the (average) magnetic
field strength by a power law with an exponent below one. Indeed, computations by \citet{1991A&A...250..220S} indicate such a saturation, although the brightness of \ion{Ca}{ii} does not follow a power law with magnetic flux, but saturates faster, quickly approaching a fixed value (for magnetic flux tubes whose temperature does not depend on flux density).

Extending this scenario, one could argue that emission
forming at higher altitudes would be more susceptible
to this effect, because at higher heights the flux
tubes will merge already at smaller average magnetic
field strengths (or less dense packing of flux tubes).
Therefore one might expect the power-law index to drop
with height from the photosphere to the chromosphere,
as we see in our observations.

However, at this point this scenario can only be a
gross speculation. Photospheric observations at high
spatial resolution of 0.15\arcsec\ to 0.18\arcsec\
show isolated kilo-Gauss
flux tubes even in the quiet Sun \citep[with a magnetic filling factor close
to unity within the spatial resolution element;][]{2010ApJ...723L.164L}. However, if we consider the visible continuum, that is, radiation coming from near the deepest observable layers, then this radiation displays a turnover with magnetic flux, except at the very highest spatial resolution, where it stays nearly flat for most of the magnetic field range \citep[e.g.][]{2017ApJS..229...12K}. Broad-band radiation at 300nm, which also clearly displays granulation (i.e. arises in the low photosphere), similarly shows a markedly nonlinear behaviour with magnetic flux. Because magnetic flux tubes are unlikely to merge in the low photosphere, it is not clear if, and if yes, to what extent the scenario of \cite{1989ApJ...337..964S}
might apply. In any case, there is a need for an alternative/additional mechanism if we consider all photospheric layers.

\subsubsection{Height of formation}\label{S:height.formation.effects}
The height of formation likely also plays a role. The observations at different wavelengths in the \ion{Mg}{ii} wings (see Fig.\,\ref{fig_s}) suggest a decrease of the power law exponent with height of formation in the upper photosphere. This raises the question why AIA\,1600\,\AA\ displays a higher power law exponent than \ion{Mg}{ii}\,k1r.

In the semi-empirical static one-dimensional model of \cite{1981ApJS...45..635V}, their Fig.\,1, the UV continuum near 1600\,\AA\ imaged by AIA and the \ion{Mg}{ii}\,k1 features should form at around the same temperature, near the temperature minimum. Therefore they should show the same statistical properties when related to the magnetic field, and this is why the significant difference in the power-law index between AIA\,1600\,\AA\ and \ion{Mg}{ii}\,k1r
comes as a surprise. However, there is evidence that the UV continua do not form near the temperature minimum but much lower in the atmosphere. Relating UV brightenings to the magnetic field,  
\cite{2016A&A...592A.100R} wondered why the magnetic concentrations are so well visible in the UV. In their discussion they argue that this ''results also from ionisation of \ion{Fe}{i} and the other neutral electron donor species since their bound-free edges dominate the UV continuous extinction in the upper photosphere.''  Consequently, the UV continuum, and also the AIA\,1600\,\AA\ channel, originate from much lower heights than expected for the \cite{1981ApJS...45..635V} type models, and by this answers the above question.

\subsubsection{Wavelength-dependent contrast}\label{S:wavelength.dependent.contrast}
Another question posed by our observations is why the power-law index might decrease with height from the source region of the 1600\,\AA\ continuum to \ion{Mg}{ii}. This could be related to the contrast being dependent on wavelength.

The contrast at 1600\,\AA\ is also expected to behave differently from that in \ion{Mg}{ii}\,k1r due to the difference in wavelength. When going to shorter wavelengths, the radiation intensity depends ever more strongly on temperature. This is well visible for thermal radiation as described by Planck's black-body equation. This means that for a slight increase in temperature there is a correspondingly stronger increase in intensity at shorter wavelengths. This in turn implies a less steep dependence of the intensity on the magnetic field for \ion{Mg}{ii}\,k1r as compared to the 1600\,\AA\ continuum and hence a drop of the power-law index from 1600\,\AA\ to \ion{Mg}{ii}\,k1r.

\subsubsection{Flux tube heating}\label{S:flux.tube.heating.effects}

An alternative explanation for the nonlinear (power-law)
relation between intensity and magnetic field, together
with its change from the photosphere into the chromosphere
might be based on the heating of magnetic flux tubes as a function of height and magnetic flux.

To understand the nonlinear relation between intensity and magnetic field in the photosphere it might be helpful to recall that  magnetic features in active regions are on
average broader than in quieter regions. In the context of our study this implies that there is the tendency for structures in regions with a high magnetic flux density (i.e. high magnetogram signal) to be broader.  This makes
them less hot and hence less bright \citep{1994A&A...285..648G},
which is also reflected in empirically derived temperatures
of magnetic flux tubes \citep{1992A&A...262L..29S}.
Consequently, the brightness per unit flux decreases
\citep{2002A&A...388.1036O, 2013A&A...550A..95Y}. Therefore we would expect a flattening of the relation between intensity and magnetic flux at higher fluxes, which is equivalent to saying that the power-law index of the relation is below one. To explain the dependence of the power law index on wavelength would imply that this heating becomes more flux dependent from the middle to the upper photosphere.

Still, at this point this set of explanations (Sect.\,\ref{S:pli.flux.merging} to \ref{S:flux.tube.heating.effects}) for our finding
of the change of the power-law index from the photosphere
to the chromosphere has to remain speculative.
To reach more solid ground, detailed radiation MHD simulations of the photosphere and chromosphere including
the formation of UV lines and continua would have to be investigated,
which is beyond the scope of this manuscript.

Understanding the cause of the power-law relation is of interest not only in the context of the present study, but also for  investigations of the general relation of the intensity in the photosphere to the magnetic field of faculae and network elements \cite[e.g.][]{2013A&A...550A..95Y}. There the relation is also highly nonlinear, at very high spatial resolution, but  is closer to a logarithmic function than to a  power law \citep{2017ApJS..229...12K}.

\subsection{Mag-flux relation from the chromosphere to the transition region}\label{S:power.law.discussion}

From the chromosphere to the transition region the power-law index characterising the relation of the emission to the surface magnetic field increases with formation temperature (Sect.\,\ref{S:cor.pow.plage.results}). Here we will turn to the question how to understand this variation of the long end of the hockey stick in Figs.\,\ref{fig_f}b, \ref{fig_h}b. This is of particular interest because the overall shape remains unchanged irrespective of the spatial resolution or region of interest (while the absolute values of the power-law indices differ). In the previous subsection, we gave two possible interpretations for the short end of the hockey stick, in other words, the drop of the power-law index from the upper photosphere to the temperature minimum. Obviously, these interpretations do not apply for the opposite variation we see through the chromosphere into the transition region.


The increase in power-law index with formation temperature basically reflects that emission forming at higher altitudes will be more sensitive to the magnetic field. Considering that the X-ray emission clearly shows a very high contrast to the magnetic field, it seems natural to find this increasing sensitivity to the magnetic field already in the chromosphere and transition region. Here the magnetic field starts dominating the plasma (plasma-$\beta$ is smaller than unity) and the magnetic heating has to take over from acoustic heating, because the latter is not sufficient to heat the higher regions of the atmosphere. This interpretation is also backed by studies of the flux-flux relations from the chromosphere to the corona of the Sun \citep{1992A&A...258..507S} and in other stars \citep{1987A&A...172..111S}. 

This provides some insight into which processes govern the relation of the emission in the upper atmosphere to the magnetic field. In their qualitative considerations \citet{1989ApJ...337..964S}
speculated in their Sect.\,VII.c that the nonlinear relation of emission and magnetic field
''can be explained as a combined effect of geometrical filling of (...) flux tubes [and] the dependence
of the heating efficiency on flux-tube packing''. In our study, where we also investigate the temperature dependence of this nonlinearity we can be more specific. To be consistent with the hockey-stick type temperature variation, from the upper photosphere into the temperature minimum region the geometrical effects (discussed in Sect.\,\ref{S:aia1600.results}) would dominate, while from the chromosphere into the transition region and probably the corona the increasing sensitivity of the emission to the magnetic heating processes becomes dominant.

\subsection{Mag-flux relation at different resolutions and regions}\label{S:mag.flux.diff.regions}

The results discussed above for the active region set at 1.5\arcsec\ resolution basically holds also for the other regions of interest and at different spatial resolutions. In the following we give some explanations where the mag-flux relations differ from the 1.5\arcsec\ active region set.  

The slight increase of the power-law indices for lower spatial resolution (cf.\ Fig.\,\ref{fig_f}b) most probably is a consequence of the basal flux. When degrading the data spatially we find that the basal flux increases slightly. This is because after the degradation, regions of formerly high magnetic field (and thus high emission) will have lower magnetic flux densities (in particular if close-by opposite polarities cancel). These areas are then included in the mask to calculate the basal flux (cf.\ Sect.\,\ref{S:basal.flux}) resulting in a larger basal flux.
Figuratively speaking, subtracting a larger basal flux pulls down the emission at low magnetic fields more strongly (in the double-logarithmic plot) and thus the power-law fit gives a steeper slope, viz.\ larger power-law index.

Of the regions under investigation, the plage region sticks out in terms of the overall level of the power-law index: the hockey stick for the plage region is well below the other areas (Sect.\,\ref{S:only.plage.results}, Fig.\,\ref{fig_h}b). This is mainly an artefact of the definition of the plage region. Because this excludes the darker quiet Sun parts (cf.\ Sect.\,\ref{S:roi}.2), here we basically cut the lower-intensity region of the scatter plot of the intensity vs. magnetic field scatter plots (e.g.\ Fig.\,\ref{fig_c}).
Consequently, the variation over the considered range of magnetic field values will be smaller and the derived power-law indices will be smaller.

For the quiet Sun and coronal hole regions the situation is different. While the correlations for these two regions are smaller than for the active region (just as the plage), the power-law indices are actually comparable (cf.\ Fig.\,\ref{fig_h}).
However, unlike in the plage regions, for the quiet Sun and coronal hole regions there is no cut-off at low intensities and hence the power-law indices are comparable to the active region.

\subsection{Flux-flux relations: the Sun and stars}\label{S:flux.flux.discussion}

Relating each emission feature to the emission forming at the lowest temperature (\ion{Mg}{ii}\,k1r) we find that the power law-index increases with formation temperature (cf.\ Table\,\ref{table:t3}, Sect.\,\ref{S:method-flux}).
Of course, this  basically reflects the relations between (unsigned) magnetic field and emission as discussed above.
Our finding is also consistent with \citet{1984NASCP2349..437B}, \citet{1989A&A...213..226C}
and \citet{1992A&A...258..507S}, all of whom noticed that the emission from the
transition region grows more rapidly  than that from the chromosphere. 

The steepening of the flux-flux relations with temperature we find here for spatially resolved data from the Sun is also found when using stellar data. For example, analysing a number of G stars \citet{1981ApJ...247..545A} showed that the flux-flux
relation has a steeper gradient going from the chromosphere into the
corona. In the stellar case some evidence has been presented that there is a clear distinction between the chromosphere and the transition region in terms of flux-flux relations. In particular, \citet{1986A&A...154..185O} found a separation of the power-law index
of the flux-flux relations between the chromospheric and the transition region
lines for F-G type stars: while 
the power-law index of \ion{the chromospheric Si}{ii} line to \ion{Mg}{ii} is only 1.2, there is a jump to the transition region lines such as \ion{C}{ii} or \ion{Si}{iv}  with a power-law index of about 1.6 \cite[][Fig.\,7]{1986A&A...154..185O}. However, our analysis does not show this jump (or separation)
we find a continuous change with formation temperature, in fact, we find only a small difference between \ion{Mg}{ii}\,k3 and \ion{C}{ii}. Maybe   
\citet{1986A&A...154..185O} found this jump only because he was not investigating the different features of the \ion{Mg}{ii} line as we do here. Also, considering the physical scenario to understand the steepening of the power-law indices from the chromosphere into the corona discussed above (Sect.\,\ref{S:power.law.discussion}) would not provide an explanation for a jump. So if this jump in stellar observations is real, the reason for this would have to be found in some global properties of the atmosphere when integrating over the whole stellar disc, for example, through differences in the centre-to-limb variation between lines forming under optically thick or thin conditions. In particular, the reason for the jump cannot be due to individual processes governing the physics of the individual magnetic flux tubes connecting the photosphere to the upper atmosphere.

While the power-law indices we derived  mostly do not depend on spatial resolution (cf.\ Sect.\,\ref{S:mag.flux.diff.regions}), there is one noticeable exception: the flux-flux relation of \ion{Si}{iv} to \ion{Mg}{ii}\,k1r. Here we see higher power-law indices for higher resolution (Fig.\,\ref{fig_g}). At this point we can only speculate on this, because further observational confirmation would be needed for this finding. It might well be that at smaller spatial scales \ion{Si}{iv} becomes more dependent on the magnetic field. This could point to the (expected) feature that the magnetic heating operates at the smallest scales we can observe, and probably even below that. 

Comparing different flux-flux relations we find that the scatter in the emission (in the flux-flux relation) has to be independent of the emission itself (Sect.\,\ref{S:flux.flux.results}).
This could be consistent with a scenario where the underlying magnetic field governs the upper atmosphere in independent ways. For example, different spatial scales of the (opposite directed) magnetic flux concentrations in the photosphere
might lead to an energisation of different atmospheric heights, hence temperature  regimes. So one would statistically expect that the chromosphere and the transition region brighten at higher magnetic field strengths, but the details, that is, if the chromospheric or transition region plasma gets brighter, might change from case to case, which then causes the considerable scatter (being independent of the actual intensity). This would also be consistent with the analysis of times series data comparing the chromosphere (UV continuum) to the lower (\ion{C}{ii}) and upper transition region (\ion{O}{vi}) as reported by \cite{2003A&A...406..363B}. Here sometimes the brightenings in all three spectral features are co-temporal, sometimes only two features brighten up, sometimes the brightening in one feature comes alone: no particular pattern could be established observationally.

\section{Conclusion}
We investigated the connection between emission originating from different temperatures from the upper solar atmosphere to each other and to the underlying magnetic field. The IRIS spectroscopic maps provide not only  unprecedented spatial resolution, but more importantly a continuous coverage from the photosphere through the chromosphere into the transition region to the corona. The wings of \ion{Mg}{ii}, the spectral features of the \ion{Mg}{ii} line core, and \ion{C}{ii} and \ion{Si}{iv} are observed with the same spectrograph which eliminates any issues of alignment. 

As expected for a magnetic field expanding into the upper atmosphere we confirm the continuous decrease of the correlation coefficient between emission and photospheric magnetic field with increasing line formation temperature. As a new result we found that this continuous decrease is also present through the spectral features of the \ion{Mg}{ii} line, that is, from k1r through k2r to k3 that form at increasing temperature  through the chromosphere according to semi-empirical (static) models.
Considering the highly dynamic and complex structure of the chromosphere \cite[e.g.][]{2009SSRv..144..317W} this is surprising.

The main observational result of this study is the change of the power-law index of the relation between the emission and the magnetic field that follows a power law. Here we find a hokey-stick like variation. From the photosphere to the temperature minimum the power-law index drops quickly and then rises again through the chromosphere into the transition region (Figs.\,\ref{fig_f}b and \ref{fig_h}b).

For the decrease of the power-law index below the temperature minimum we speculated about the role of four effects. These are related to the expansion of magnetic flux tubes (Sect.\,\ref{S:pli.flux.merging}), and to the location of the source region of the radiation (Sect.\,\ref{S:height.formation.effects}). The dependence of the intensity-temperature relationship on wavelength (Sect.\,\ref{S:wavelength.dependent.contrast}), and the dependence of the heating of flux tubes on the magnetic flux density (Sect.\,\ref{S:flux.tube.heating.effects}) likely also play a role in determining the observed behaviour. Clearly, further work is needed to draw final conclusions on the role of these and possibly other effects. The increase
of the power-law index with temperature in the chromosphere
and transition region most probably is related to the sensitivity
of the emission from the upper atmosphere to the magnetic
heating process, which becomes even more sensitive when going further into the high-temperature regimes of the corona.

\begin{acknowledgements}
      We would like to thank to Rob Rutten for many constructive comments and Robert Cameron, Leping Li, Chen Nai-Hwa, and Holly Waller for helpful suggestions. 
      
      This work was supported by the International Max-Planck Research School (IMPRS) for Solar System Science at the University of G\"ottingen.
      
      L.P.C. received funding from the European Union's Horizon 2020 research and innovation programme under the Marie Sk\l{}odowska-Curie grant agreement No.\,707837 and acknowledges previous funding by the Max-Planck-Princeton Center for Plasma Physics. 
      
      This project has received funding from the European Research Council (ERC) under the European Union's Horizon 2020 research and innovation programme (grant agreement No. 695075) and has been supported by the BK21 plus programme through the National Research Foundation (NRF) funded by the Ministry of Education of Korea.
\end{acknowledgements}

\bibliographystyle{aa}
\bibliography{flux_flux}

\clearpage


\begin{appendix}

\section{Basal flux calculation} \label{S:bflux}

\begin{figure*}
\begin{center}
\resizebox{0.495\hsize}{!}{\includegraphics{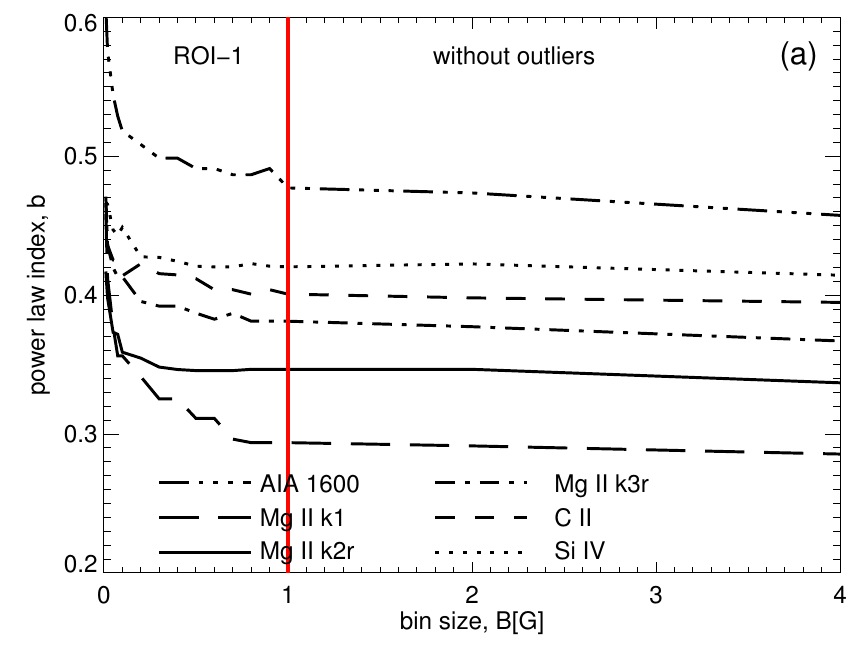}} 
\resizebox{0.495\hsize}{!}{\includegraphics{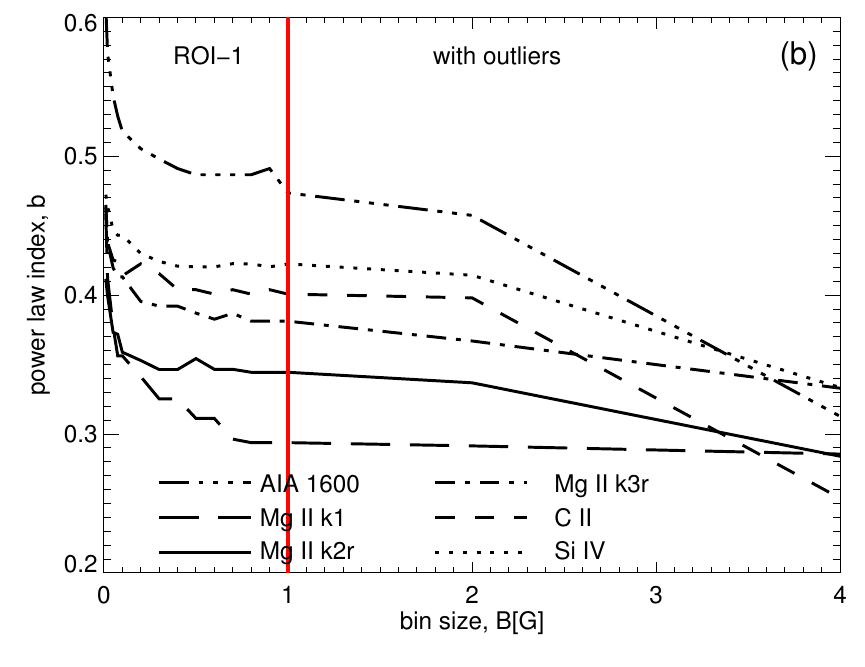}} 
\caption{Power-law relation and basal flux.
The left panel (a) presents the power-law-index $b$ between magnetic field and the emission flux proxies for different values of the basal flux determined in bins of different sizes (for all data points with magnetic field strengths below 4\,G).
For larger bin-sizes the basal flux, and thus the derived power law index is constant.
The right panel (b) shows the same, but for data sets where we added outliers to the data. 
The vertical red lines indicates the bin size we use in our study.
See Appendix\,\ref{S:bflux}. \label{fig_i}}
\end{center}
\end{figure*}

To investigate the relation between the magnetic field strength, $|B|$, and the flux in an emission line or continuum, $I$, one has to determine the basal flux. In Sect.\,\ref{S:basal.flux} we briefly discussed the method we employ to calculate the basal flux. In this appendix we give some further details on the basal flux determination and in particular discuss what impact this has on the power law index characterising the relation between $|B|$ and $I$.

As outlined in Sect.\,\ref{S:basal.flux}, in their original work \citet{1989ApJ...337..964S} defined the basal flux as the component of emission which originates independently of the magnetic field from acoustic processes, for example, the heating of the chromosphere through the dissipation of sound waves that steepen into shocks.
To determine the basal flux we concentrate on the nonmagnetic regions in the respective region-of-interest, in other words, we consider only spatial pixels where the signal of the magnetogram is consistent with noise.
In our study we use HMI which has a noise level of about 5-8\,G \citep{2013A&A...550A..95Y}. To be on the safe side concentrating on nearly-field-free regions, we use only half of the noise level as the upper limit for the magnetic field strength, that is, 4\,G.

Because of the limited spatial resolution and polarimetric sensitivity these nearly-field-free regions still might host significant magnetic patches, simply hidden to the instrument \cite[e.g.][]{2017ApJS..229....4C}.
Thus one cannot simply take, for example, the mean or median value of the intensity distribution in these areas with low magnetic field (below 4\,G) --- this would give a too high value of the basal flux because most probably one would include quite a few regions with hidden magnetic field that boost the emission.
Instead one should still follow the concept of \citet{1989ApJ...337..964S} and find the minimum value of the emission.

If one would simply define the minimum of the intensity in all the regions with a magnetic field strength below a threshold (4\,G) one would be sensitive to outliers. For example, if there are only a few pixels with very small intensity values (for whatever reason) these would result in a gross underestimation of the basal flux.
To avoid this problem of outliers one could consider to define the basal flux based on the distribution of the intensity in these regions of low magnetic signal (below 4\,G). For example, one could define the basal flux at the fifth or tenth percentile level of the distribution. Or one could use the width of the distribution of intensities to define the basal flux, for example, the mean value minus the standard deviation of the intensities.
However these definitions rely on the bulk part of the distribution of intensity values. Thus they are not well suited to define the basal flux, that is, a minimum value of intensities. Still, we tested this, but as expected we found that the basal flux defined through a percentile or the standard deviation would be very sensitive to the choice of parameters (e.g. fifth or tenth percentile, one or two standard deviations). Therefore these strategies are not a good procedure to determine the basal flux without sensitivity to outliers.

To overcome the sensitivity to outliers we subdivide the regions below a threshold (of 4\,G) into bins with respect to the magnetic field. We choose to have bins linearly spaced in magnetic field strength. In each of these bins we determine the minimum intensity and then define the median value of these minimum values in the bins as the basal flux.

To test our method and to determine the optimum size for the bins in magnetic field strength, we calculate the basal flux for a number of different bin sizes. In the end we want to get reliable results for the power law index of the relation between the magnetic field and intensity (Sect.\,\ref{S:power.law.fit}). Thus we used the basal fluxes determined from our procedure with different bin sizes, corrected the intensities for the respective basal flux and calculate the power law index for each of the bin sizes.
The results are shown in Fig.\,\ref{fig_i}. We concentrate first on the left panel that shows a data set without outliers.
As expected for small bin sizes we get too large basal fluxes that result in steeper power laws (i.e. larger power law indices $b$). Basically, when the bin size is too small there will be only a few data points in each bin and thus, statistically speaking, the mean minimum intensity in each bin will not be too different from the mean in the same bin. Then the median of the minima in the bins, that is, the basal flux as we define it, will be close to the mean value of the distribution of all intensities below the magnetic field threshold (of 4\,G). Consequently too small bin sizes overestimate the basal flux which also leads to too large power-law-indices.
From the left panel of Fig.\,\ref{fig_i} we conclude that the bin size should be about 0.7\,G or larger. For larger bin sizes the derived power-law-indices for the different emission proxies remain roughly constant.

To check that our method works also in the presence of outliers we took one data set without outliers and added extra data points to each analysed intensity channel with low intensity to mimic outliers.
We apply the same procedure as before and calculate the power-law index as a function of bin size for different emission proxies and show the result in the right panel of Fig.\,\ref{fig_i}.
The largest bin size shown in the plot, 4\,G, implies that there is only one bin, that is, the basal flux is the minimum of all pixels below the threshold of magnetic field strength, just as in the original method by \citet{1989ApJ...337..964S}.
We see that for too large bins the basal flux and hence the power-law-index becomes sensitive to the outliers resulting in too small power-law indices.
Still, this effect becomes significant only for bin sizes above about 2\,G.
Of course, the exact values of the optimum bin sizes will depend on the data set used and the number of data points considered in the analysis.

For our data set we find that with a bin size of about 1\,G we get reliable and robust results for the power-law-index characterising the relation between magnetic field and photon flux in different emission proxies, that is, the AIA 1600\,\AA\ continuum channel, the features of the chromospheric \ion{Mg}{ii} line, and the transition region lines of \ion{C}{ii} and \ion{Si}{iv}.
Our method extends the original procedure by \citet{1989ApJ...337..964S} allowing us to cope with outliers and at the same time avoids the sensitivity to the choice of parameters when using percentiles of the intensity distribution or the standard deviation of the intensities to calculate the basal flux.

\newpage

\section{Comparing different methods to determine power-law relations}\label{S:pl-methods}

\begin{figure*}
\sidecaption
\includegraphics[width=12cm]{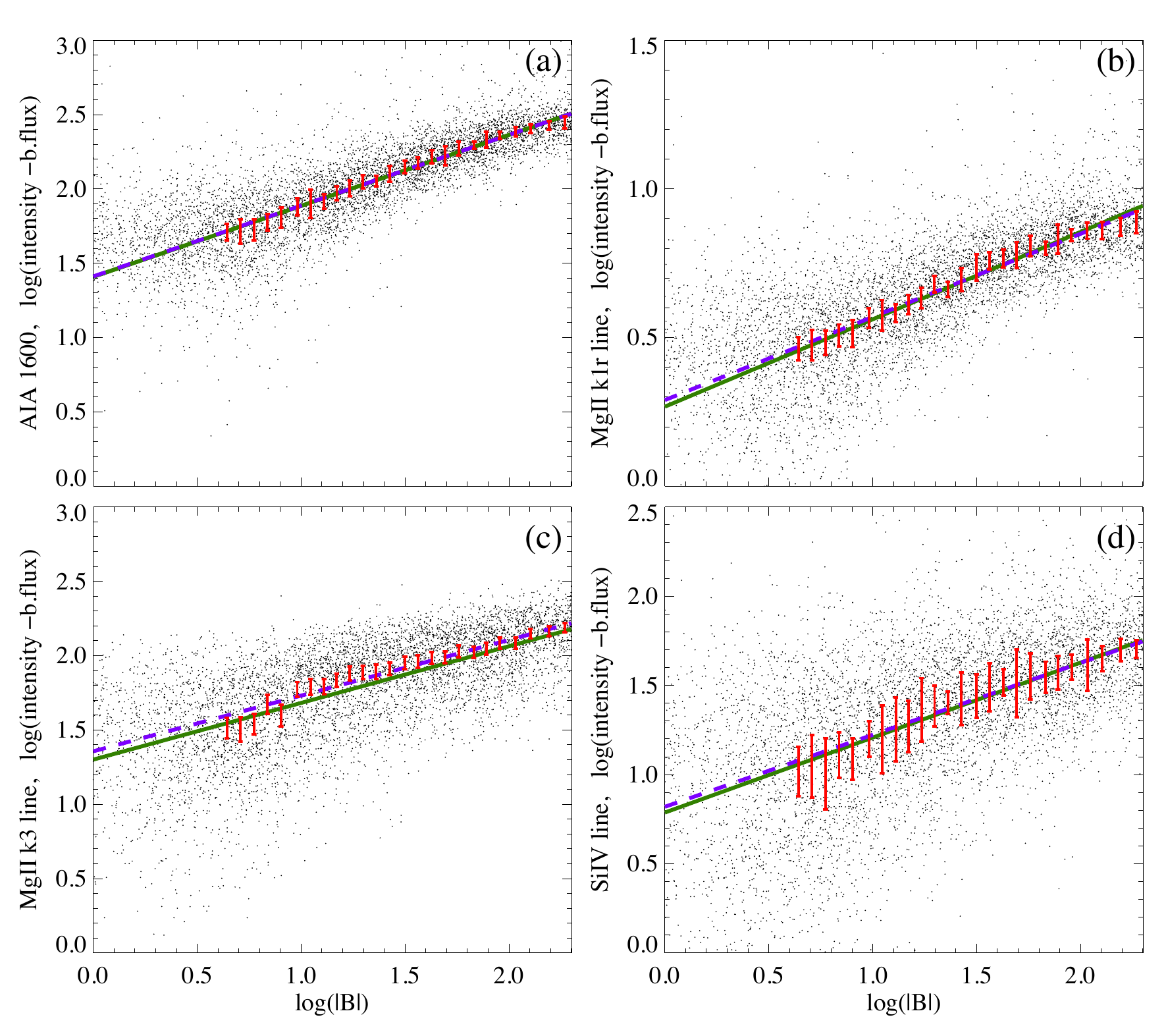}
\caption{Relation of the upper atmosphere emission to the underlying magnetic
field. This shows the same data as Fig.\,\ref{fig_c}, but now on a double-logarithmic
scale to emphasise the power-law nature of the relations. Here the basal flux (b.flux) in the respective emission feature is subtracted. See Sect.\,\ref{S:power.law.fit}
and Appendix\,\ref{S:pl-methods}.
\label{F:BI.log.log}}
\end{figure*}

In Sect.\,\ref{S:power.law.fit} we discussed two methods (below labelled I and II) to determine the power-law index $b$ of the nonlinear relation between emission proxies and the underlying magnetic field. There we imaged the results in a linear plot in Fig.\,\ref{fig_c}. Here we show the same data but on a double-logarithmic scale in Fig.\,\ref{F:BI.log.log}. 

To check the reliability of the obtained power-law fits, we used several methods and applied them to the same data points. In the following we recall the two methods discussed in 
Sect.\,\ref{S:power.law.fit} (I and II) and briefly explain three further methods (III--V) to derive the power-law indices. 

(I)~
Least-squares fit to  original data points. After subtracting the basal flux $I_0$ from the intensities $I$ giving $I^\prime{=}I{-}I_0$ (Appendix\,\ref{S:bflux}) we fit a power law, $I^\prime=a\,|B|^b$. 

(II)~ 
Least-squares fit to binned data. We first collect the data points into bins with respect to the magnetic field strength and compute the average value of the intensity in each bin (bars in Fig.\,\ref{fig_c}) and then perform the power-law fit to these averages.

\begin{figure}
   \centering
   \includegraphics[width=8.8 cm]{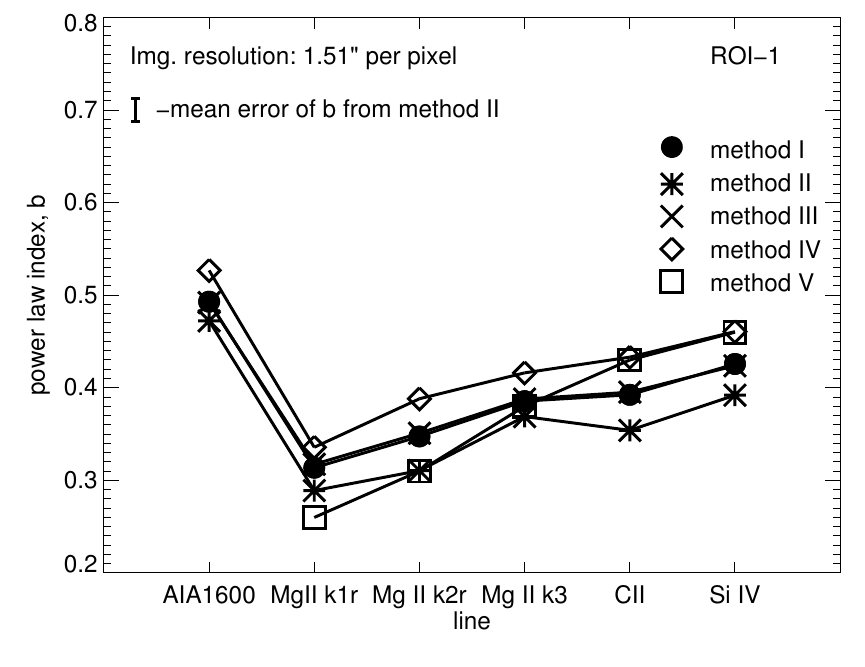} 
   \caption{Comparison of power-law indices derived from five methods.
This plot shows the power-law indices of the relation of intensity vs.\ magnetic field for the various emission features in the same way as displayed in    Figs.\,\ref{fig_f}b and \ref{fig_h}b. The five methods are briefly described in Appendix\,\ref{S:pl-methods}.} 
              \label{fig_j}%
  \end{figure}

(III)~
Variant of Method I with different basal flux.
As discussed in Appendix\,\ref{S:bflux}, to compute the basal flux $I_0$ we take the median value of the minimum intensities  in bins below a magnetic field threshold (4\,G). To go to the extremes of basal fluxes, we also take the minimum, $I_0^{\rm{(min)}}$, and the maximum values, $I_0^{\rm{(max)}}$, of the intensities below the threshold. For these three basal fluxes we perform the power-law fit and derive three (different) values for the power-law index. In Fig.\,\ref{fig_j} we plot the mean value of the three indices.

(IV)~
Linear fit to  logarithmic data values. This is similar to method I, but using a linear least-squares fit to the data points on a logarithmic scale, that is, $(\log I^\prime)= b\;(\log|B|)+\tilde{a}$.
This should be equivalent to the power-law fit to the original data, of course, with the slope of the linear fit taking the role of the power-law index.

(V)~
Ellipse fitting of the logarithmic data values.
In this method we fit ellipses to the (2D) probability distribution functions, that is, the point density of the scatter plots $\log I^\prime$ vs.\ $\log|B|,$
as shown in Fig.\,\ref{F:BI.log.log}.
This follows the procedure as described in Sect.\,\ref{S:method-flux} to derive the power-law indices of the flux-flux relations.

These methods agree well with respect to the power-law indices. In Fig.\,\ref{fig_j} we show the power-law indices derived by the five methods for the various emission features ordered by formation temperature.
In particular all the five methods show the hockey-stick feature (cf.\ Figs.\,\ref{fig_f}b and \ref{fig_h}b; Sect.\,\ref{S:cor.pow.plage.results}). Therefore this feature  does not depend on the method employed and the discussion on the physical implications of this in Sects.\,\ref{S:aia1600.results} and \ref{S:power.law.discussion} is based on a solid observational footing.

\end{appendix}

\end{document}